\documentclass{aa}
\usepackage{graphicx}
\begin{document}
    \title{Astrometric radial velocities}
    \subtitle{III. Hipparcos measurements of nearby star clusters and 
associations
\thanks{Based on observations by the ESA Hipparcos satellite.
Extended versions of Tables~\ref{tab1} and \ref{tab2} 
are available in electronic form at the CDS via anonymous 
ftp to cdsarc.u-strasbg.fr}
}
    \author{S{\o}ren Madsen, Dainis Dravins, and Lennart Lindegren}
    \offprints{S. Madsen, {\tt soren@astro.lu.se}}
    \institute{
    Lund Observatory, Box~43,
    SE--22100 Lund, Sweden\\
    \email{soren, dainis, lennart@astro.lu.se}}

\date{Received -- -- --; accepted -- -- --}
\titlerunning{Astrometric radial velocities III.}
\authorrunning{S{\o}ren Madsen et al.}

\abstract{
Radial motions of stars in nearby moving clusters are determined from 
accurate proper motions and trigonometric parallaxes, without any use 
of spectroscopy.  Assuming that cluster members share the same velocity 
vector (apart from a random dispersion), we apply a maximum-likelihood 
method on astrometric data from Hipparcos to compute radial and space 
velocities (and their dispersions) in the Ursa Major, Hyades, Coma 
Berenices, Pleiades, and Praesepe clusters, and for the 
Scorpius-Centaurus, $\alpha$~Persei, and `HIP~98321' associations.   
The radial motion 
of the Hyades cluster is determined to within 0.47~km~s$^{-1}$ (standard 
error), and that of its individual stars to within 0.6~km~s$^{-1}$.
For other clusters, 
Hipparcos data yield astrometric radial velocities with typical accuracies 
of a few km~s$^{-1}$.  
A comparison of these astrometric values with spectroscopic radial 
velocities in the literature shows a good general agreement and, 
in the case of the best-determined Hyades cluster, also permits searches
for subtle astrophysical differences, such as evidence for enhanced
convective blueshifts of F-dwarf spectra, and decreased gravitational
redshifts in giants.
Similar comparisons for the Scorpius OB2 complex indicate some
expansion of its associations, albeit slower than expected from their ages.
As a by-product from the radial-velocity solutions, kinematically improved
parallaxes for individual stars are obtained, enabling Hertzsprung-Russell
diagrams with unprecedented accuracy in luminosity.  For the Hyades
(parallax accuracy 0.3 mas), its main sequence resembles a thin line,
possibly with wiggles in it.  Although this main sequence has underpopulated
regions at certain colours (previously suggested to be 'B{\"o}hm-Vitense gaps'),
such are not visible for other clusters, and are probably spurious.
Future space astrometry missions carry a great potential 
for absolute radial-velocity determinations, insensitive to the 
complexities of stellar spectra.
}

\maketitle


\begin{figure*}
   \resizebox{\hsize}{!}{\includegraphics*{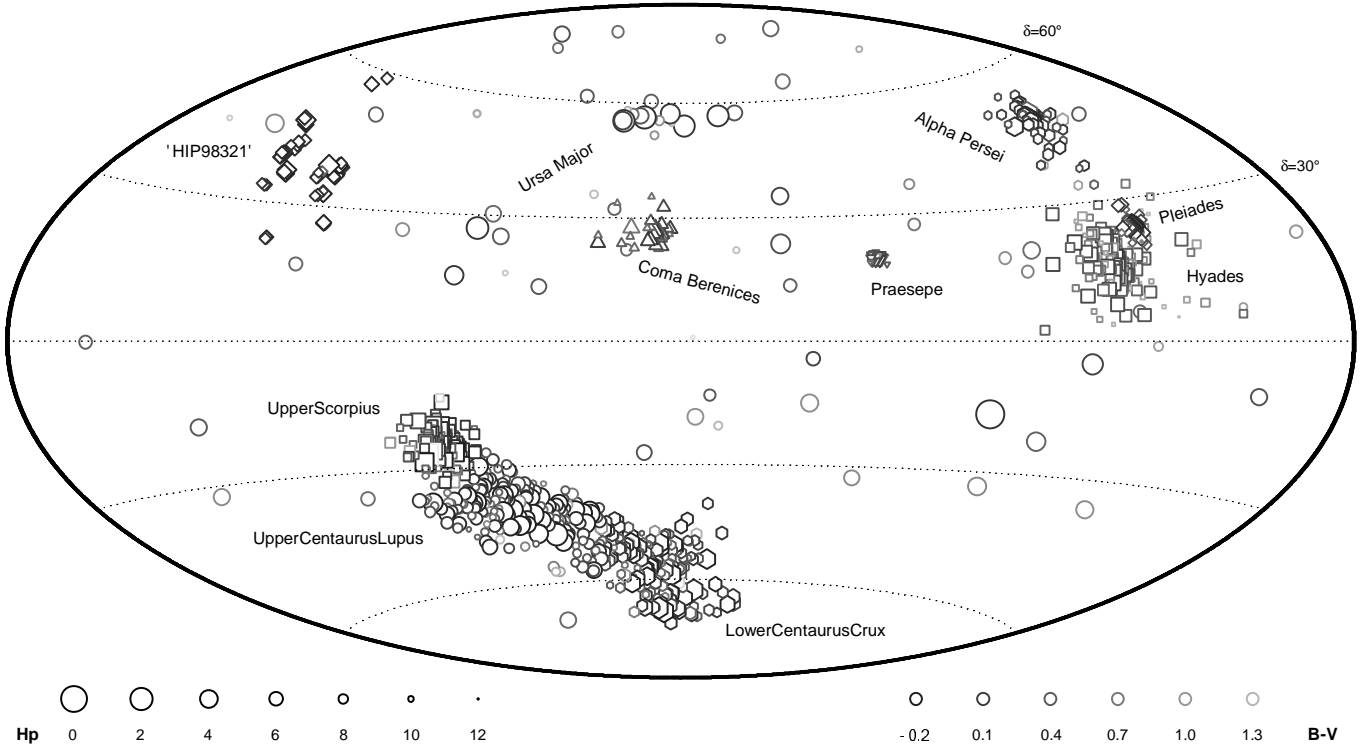}}
  \caption{Map of the full sky, showing those stars in 
clusters and associations, whose radial velocities were astrometrically 
determined from Hipparcos data. Symbol shape identifies different clusters; 
symbol size denotes apparent magnitude {\it Hp} ($\simeq m_V$), while symbol 
shading denotes $B\!-\!V$ (note how some clusters are dominated by very blue 
stars). The Aitoff projection in equatorial coordinates is used, with 
$\delta=0^\circ$ on the major axis and $\alpha=180^\circ$ on the 
minor axis. Right ascension ($\alpha$) increases to the left.}
\label{fig1}
\end{figure*}
%
%
\begin{figure*}
   \resizebox{\hsize}{!}{\includegraphics*{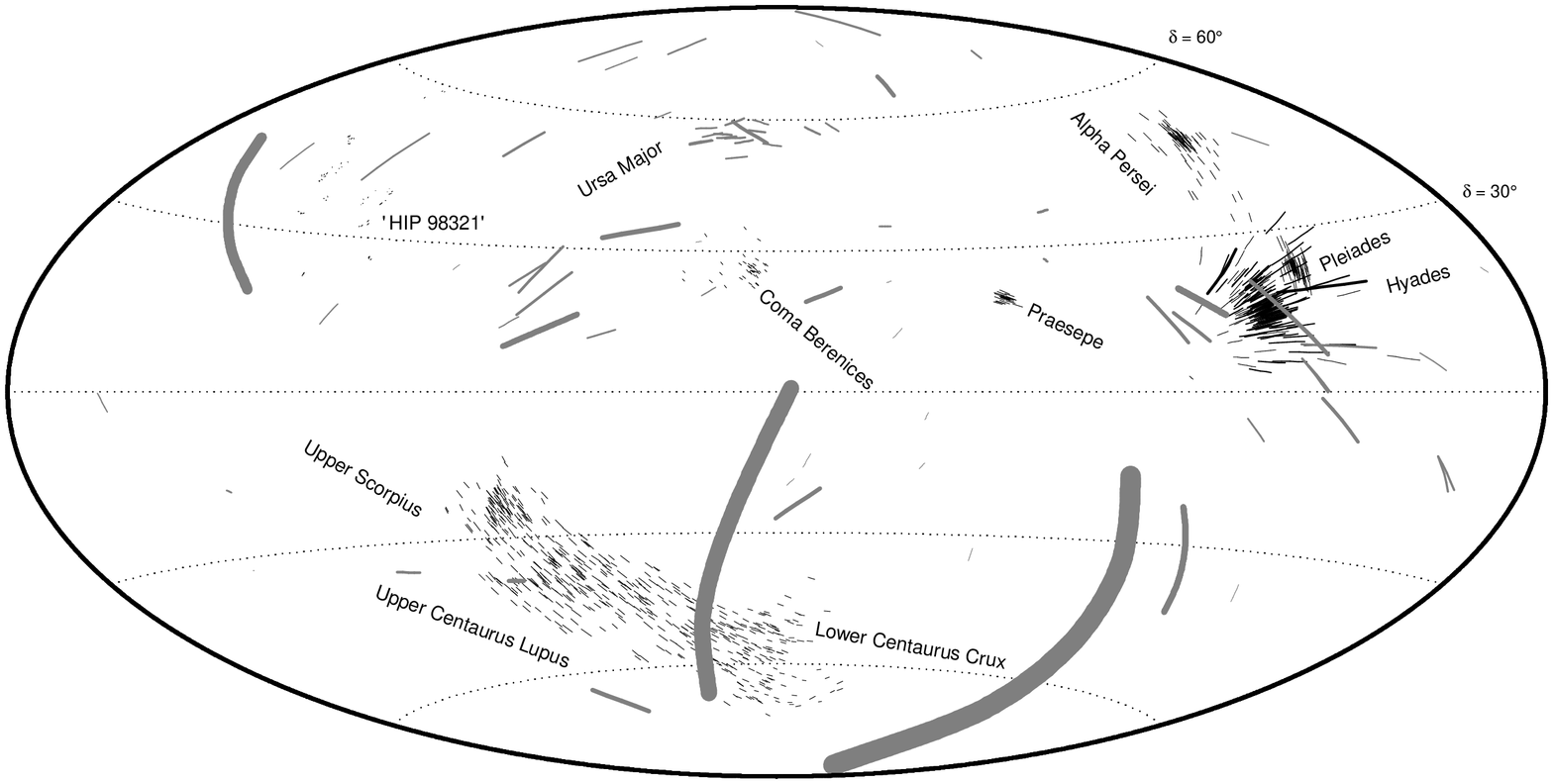}}
  \caption{Proper motions of the programme stars over 200,000 years. Best 
radial-velocity accuracy is obtained in rich nearby clusters with large 
angular extent, and large proper motions. However, the accuracy in the largest 
associations (Ursa Major, Scorpius-Centaurus) is limited by the partly unknown 
expansion of these systems. Stellar paths in the Ursa Major group (shown shaded) 
cover large areas of the sky. The thickness of the proper-motion vectors is 
inversely proportional to stellar distance: the closest star is Sirius and the 
two next ones are faint red dwarfs. Proper motions vary greatly among 
different clusters.}
\label{fig2}
\end{figure*}

\section{Introduction}

This paper is the third of a series on the determination of stellar radial 
motion by purely geometric measurements. Such {\it astrometric radial 
velocities\/} allow to disentangle stellar motion from other astrophysical 
phenomena causing spectroscopic line shifts, such as internal motions in stellar 
atmospheres and gravitational redshifts. Paper~I (Dravins et al.\ \cite{dravinsd}) 
discussed different methods which permit radial velocities to be determined 
independent of spectroscopy. Among these, the moving-cluster method of changing 
angular separation permits radial-velocity accuracies on a sub-km~s$^{-1}$ level 
to be reached already with current astrometric measurements, such as those from 
the Hipparcos space mission (ESA \cite{esa}). That method analyses the changing 
angular extent of a star cluster as it approaches or recedes from the Sun, 
assuming that the member stars share the same average velocity vector relative 
to the Sun. The radial-velocity component makes the cluster appear to expand
or contract due to its changing distance.  This relative rate of apparent 
contraction equals the relative rate of change in distance to the cluster, from 
which a linear velocity (in km~s$^{-1}$) follows whenever the absolute distance 
is known from trigonometric parallaxes. 

That it is possible, in principle, to determine radial velocities from
astrometry has been known for a long time. Attempts were made e.g.\ by
Petrie (\cite{petrie49}, \cite{petrie63}) and Eggen (\cite{eggen84}, \cite{eggen98}) to derive
radial velocities that are at least partially independent of spectroscopy.
The availability of the Hipparcos results allowed to investigate the 
kinematics and memberships of moving groups and clusters in great detail, 
which in turn made it possible to apply our moving-cluster method to reach better 
accuracies. The detailed mathematical formulation of the method was 
presented in Paper~II (Lindegren et al.\ \cite{lindegren00}).

As a by-product of the moving-cluster method, the distance estimates to 
individual cluster stars are often significantly improved compared with the 
original [here Hipparcos] parallax measurements. These {\it kinematically 
improved parallaxes\/} (Paper~II) result from a combination of trigonometric and 
kinematic distance information, where kinematic distances follow from the 
observed proper motions and the derived cluster velocity.  The improved 
distances sometimes allow the Hertzsprung-Russell diagram of a cluster to be 
studied with unprecedented resolution in absolute magnitude. 

In this paper we apply this `moving-cluster method' to several nearby open 
clusters and associations. Astrometric radial velocities and kinematically 
improved parallaxes are deduced for stars in the the Ursa Major, Hyades, Coma 
Berenices, Pleiades, and Praesepe clusters, and for the Lower Centaurus Crux,
Upper Centaurus Lupus, Upper Scorpius, $\alpha$ Persei, and `HIP~98321' 
associations.

\section{Potential of the moving-cluster method}
\label{sec:method}

The achievable accuracies of the moving-cluster 
method were discussed in Paper~I. 
We recall that the best radial-velocity accuracy is obtained for star-rich 
nearby clusters with large angular extent, large proper motions, small internal 
velocity dispersions, and small rates of cluster expansion.  The improved 
astrometric accuracies expected from future space missions will somewhat lessen 
these constraints, although the intrinsic limitations set by internal velocities
cannot be overcome by increasing observational accuracy.

Only about ten clusters and associations can be meaningfully studied already with 
current astrometric accuracies in the milliarcsecond range. Furthermore, good 
astrometric data are available only for their relatively brighter stars. The 
distribution on the sky 
and the geometries of these nearby clusters are shown in Figs.~\ref{fig1} 
and \ref{fig2}, together with their stellar populations, and their proper-motion 
patterns.  Only a few among these (Hyades, Pleiades, Coma Berenices, 
Praesepe) have a utilizable stellar population spanning many spectral types; 
several of the others are heavily dominated by early-type stars. The main reason
for this is of course the limiting magnitude of the Hipparcos mission. The areas 
subtended on the sky differ greatly: some OB-associations spread out over much of 
a hemisphere, while some other clusters are very localised.  Although the great 
spatial extents of the Ursa Major cluster and the Scorpius-Centaurus associations 
in principle are advantageous for the accuracy obtainable, the partly unknown 
expansion rates and internal velocity patterns of these younger stellar groups 
actually limit the accuracy in their radial-velocity determinations.  For more 
data on these clusters, see Table~4 in Paper~I.

\begin{table*}
  \caption[ ]{Estimated space-velocity components and internal velocity 
dispersions of clusters and associations analyzed with the moving-cluster 
method, using astrometric data from the Hipparcos main catalogue.  $n_{\rm acc}$ 
is the number of stars retained in the cleaned sample for each cluster, 
$n_{\rm rej}$ is the number of stars removed in the cleaning process; 
$\widehat{v}_{0x}$, $\widehat{v}_{0y}$, and $\widehat{v}_{0z}$ are the 
equatorial (ICRS) components of the estimated space velocity of the cluster 
centroid; $\widehat{\sigma}_v$ is the estimated internal velocity dispersion 
among individual stars, calculated as described in Paper~II, Appendix~A.4.
The last three columns give the equatorial coordinates $(\alpha_0,\delta_0)$
[deg] of the adopted centroid of the cluster (or, in the case of Ursa Major,
its core) and the space velocity component $\widehat{v}_{0r}$ toward that
direction, i.e.\ an approximate radial velocity of the cluster as a whole.  
Uncertainties are given as $\pm 1$~standard error.  All velocities are in 
km~s$^{-1}$. The electronic version of the table contains, additionally,
the six equatorial components of the formal covariance matrix
$\mbox{Cov}(\widehat{\vec{v}}_0)$, the spherical equatorial coordinates 
of the convergence point with standard errors, and the total 
velocity with its standard error.}
\begin{tabular}{l*{9}{@{\extracolsep{9pt}}r}} 
  \hline\noalign{\smallskip}
   Name & 
   $n_{\rm acc}$ & 
   $n_{\rm rej}$ & 
   \multicolumn{1}{c}{$\widehat{v}_{0x}$} & 
   \multicolumn{1}{c}{$\widehat{v}_{0y}$} & 
   \multicolumn{1}{c}{$\widehat{v}_{0z}$} &
   \multicolumn{1}{c}{$\widehat{\sigma}_v$} &
   \multicolumn{1}{c}{$\alpha_0$} &
   \multicolumn{1}{c}{$\delta_0$} &
   \multicolumn{1}{c}{$\widehat{v}_{0r}$} \\
  \noalign{\smallskip}\hline\noalign{\smallskip}
  \noalign{\smallskip}
Ursa Major             &  77&  4&$ +8.44\pm 0.41$&$ -12.19\pm 0.39$&$ -10.16\pm 0.43$&$ 2.82\pm 0.23$&$ 187.3$&$ +56.4$&$ -12.24\pm 0.46$\\
Hyades                 & 168& 29&$ -5.90\pm 0.13$&$ +45.65\pm 0.34$&$  +5.56\pm 0.10$&$ 0.49\pm 0.04$&$  66.5$&$ +16.9$&$ +39.42\pm 0.36$\\
Coma Berenices         &  40&  0&$ -0.82\pm 0.96$&$  +4.57\pm 0.15$&$  -4.11\pm 0.48$&$ 0.47\pm 0.09$&$ 187.5$&$ +26.4$&$  -1.64\pm 1.07$\\
Pleiades               &  60&  0&$ +1.99\pm 2.20$&$ +22.95\pm 3.34$&$ -18.73\pm 1.82$&$ 0.50\pm 0.13$&$  56.4$&$ +24.0$&$ +10.85\pm 4.36$\\
Praesepe               &  24&  0&$ -1.46\pm 9.03$&$  +48.1\pm 10.9$&$  +2.00\pm 5.02$&$ 0.67\pm 0.23$&$ 130.2$&$ +19.6$&$  +36.2\pm 15.0$\\
Lower Cen Crux         & 179&  1&$ -0.94\pm 0.29$&$ +18.36\pm 0.15$&$  -8.59\pm 0.46$&$ 1.13\pm 0.07$&$ 189.6$&$ -56.2$&$  +5.95\pm 0.53$\\
Upper Cen Lupus        & 218&  3&$ -4.01\pm 0.27$&$ +16.47\pm 0.32$&$ -12.91\pm 0.37$&$ 1.23\pm 0.08$&$ 230.3$&$ -41.6$&$  +1.01\pm 0.51$\\
Upper Scorpius         & 120&  0&$ -3.73\pm 0.56$&$  +9.36\pm 1.09$&$ -14.57\pm 0.57$&$ 1.33\pm 0.12$&$ 243.4$&$ -24.1$&$  -0.17\pm 1.33$\\
Sco OB2                & 510& 11&$ -1.72\pm 0.15$&$ +18.19\pm 0.15$&$ -10.43\pm 0.21$&$ 1.52\pm 0.06$&$ 225.1$&$ -43.9$&$  -1.17\pm 0.26$\\
$\alpha$ Per (Per OB3) &  78&  1&$ -3.23\pm 0.89$&$ +27.15\pm 1.26$&$ -11.76\pm 1.64$&$ 0.71\pm 0.13$&$  52.9$&$ +47.8$&$  +4.53\pm 2.18$\\
`HIP 98321'            &  59&  0&$ -3.45\pm 0.66$&$ +15.55\pm 1.26$&$ -12.27\pm 1.13$&$ 2.56\pm 0.26$&$ 297.5$&$ +39.4$&$ -19.68\pm 1.74$\\
  \noalign{\smallskip}\hline
  \end{tabular}
  \label{tab1}
\end{table*}

\section{Exploitation of the moving-cluster method}
\label{sec:estim}

\subsection{Basic cluster model}
\label{sec:basic}

The mathematical procedure described in Paper~II yields maximum-likelihood 
estimates for 
the space velocity of the cluster centroid $\widehat{\vec{v}}_0$, for the 
internal velocity dispersion of stars within the cluster $\widehat{\sigma}_v$, 
and for the kinematically improved parallax of each star $\widehat{\pi}_i$,
$i=1,\,2,\,\dots\,n$. (We reserve index $0$ for the centroid. The caret 
$\widehat{\phantom{x}}$ signifies an estimated value.)
Although additional model parameters could be included, e.g.\ to describe a
possible rotation or non-isotropic dilation of the cluster, the present
studies are restricted to the `basic cluster model' in which no such systematic
velocity patterns are assumed. In Paper~II it was shown, through Monte Carlo
simulations of the Hipparcos observations of the Hyades cluster, that the
presence of any reasonable amount of rotation and shear in the actual cluster 
will not significantly bias the solution for the centroid velocity, even 
though the analysis is restricted to the basic model. One important exception 
concerns the possible isotropic expansions of gravitationally unbound 
associations. These could indeed introduce significant biases, which are
discussed separately in Sect.~\ref{sec:expand}.

\subsection{Observational data}
\label{sec:obs}

The Hipparcos Catalogue (ESA \cite{esa}) provided input data for each star $i$ 
in the form of positions in barycentric right ascension ${\alpha}_i$, 
declination ${\delta}_i$, trigonometric parallax $\widetilde{\pi}_i$, 
the proper-motion components $\widetilde{\mu}_{\alpha i}$ and 
$\widetilde{\mu}_{\delta i}$, and standard deviations and correlation 
coefficients for the latter three. (The tilde $\widetilde{\phantom{x}}$ 
signifies an observed value, the uncertainty of which needs to be taken 
into account in the estimation procedure.) The positions and proper motions
in the Hipparcos Catalogue are referred to the barycentric ICRS reference 
system, and consequently all resulting velocities are also in that system. 

For each cluster or association, an initial sample of probable member stars was 
identified from the literature, mainly from studies based on Hipparcos data.  
However, since the mathematical formalism for 
obtaining radial velocities is strictly applicable only to stars sharing the same 
average velocity vector (with a random spread about that value), cluster non-members 
and binary stars in non-modelled orbits must be removed from the sample as far 
as possible.  This was done using an iterative rejection procedure described in 
Paper~II.  Monte Carlo simulations (Sect.~4.2 in that paper) showed that this
procedure works best when adopting the goodness-of-fit rejection limit 
$g_{\rm lim}=15$.  As illustrated there for the Hyades, this gave the lowest 
scatter in the centroid radial velocity, as well as in other quality 
indicators.  Therefore, unless otherwise stated, all samples discussed here were 
cleaned according to this criterion.

Solutions were primarily obtained using data from the main Hipparcos catalogue 
but, for some clusters, we used also data from the Tycho-2 catalogue (H{\o}g 
et al.\ \cite{hog}).  The accuracies in the latter are generally somewhat 
worse, but since its proper motions incorporate about a century of ground-based 
observations, the segregation of long-period binaries may be improved.

\subsection{Clusters and associations studied}
\label{sec:clusstud}

Table~4 in Paper~I lists some 15 clusters and associations, for 
which Hipparcos-type accuracies could potentially yield viable 
astrometric radial velocities, i.e.\ with standard errors less than 
a few km~s$^{-1}$. In practice the resulting accuracies depend on
several additional factors, not considered in that survey, such as 
the number of member stars actually observed by Hipparcos, the 
position of the cluster on the sky, the statistical correlations 
among the astrometric data, and the procedures used to obtain a
clean sample. The simplest way to find out whether our method
`works' on a particular group of stars is to make a trial solution. 
We have done that for all the potentially interesting clusters and
associations, based on the list in Paper~I supplemented with data
from the compilations in the Hipparcos Input Catalogue (Turon et 
al.\ \cite{turon}), by Robichon et al.\ (\cite{robichon}), and by de Zeeuw et al.\ 
(\cite{dezeeuw}).

The original criterion for including a cluster or association in the
present study was that it yielded a valid solution with the basic
cluster model (Sect.~\ref{sec:basic}), including a non-zero estimate
of the velocity dispersion. This turned out to be the case for four
clusters (Ursa Major, Hyades, Coma Berenices, and Praesepe) and four
associations (Lower Centaurus Crux, Upper Centaurus Lupus, Upper 
Scorpius, and `HIP~98321'). For two more, the Pleiades cluster and 
the $\alpha$~Persei association, reasonable solutions could be 
obtained by assuming zero velocity dispersion in the maximum-likelihood 
procedure (note that the dispersion could still be estimated from the 
proper-motion residuals, as explained in Sect.~\ref{sec:bias}). Considering the 
astrophysical importance of these clusters, they were therefore 
included in the study. A separate solution was also made for the 
Sco~OB2 complex (Sect.~\ref{sec:sco}).

\subsection{Mathematical bias and noise in the solutions}
\label{sec:bias}

Resulting space velocities and internal velocity dispersions for the
11 clusters and associations are in Table~\ref{tab1}. $\widehat{v}_{0x}$, 
$\widehat{v}_{0y}$, and $\widehat{v}_{0z}$ are the equatorial components (ICRS 
coordinates) of the estimated space velocity of the cluster centroid,
while $\widehat{\sigma}_v$ is the velocity dispersion in each coordinate,
i.e.\ the standard deviation of peculiar velocities along a single axis.

The maximum-likelihood estimation tends to underestimate the velocity dispersion, 
as examined through Monte Carlo simulations in Paper~II.  In Appendix~A.4 of that
paper we gave an alternative procedure to estimate the velocity dispersion from 
the proper-motion residuals perpendicular to the centroid velocity projected on 
the sky. This was shown to give nearly unbiased results. 
The velocity dispersions $\widehat{\sigma}_v$ in Table~\ref{tab1} and elsewhere in 
this paper have therefore been estimated through this alternative procedure.

As was also described in Paper~II, the radial-velocity errors among individual 
stars in the same cluster are not statistically independent, but may carry a 
significant positive correlation.  For each star, the error contains a [nearly 
constant] component being the uncertainty in the cluster velocity as a whole, 
plus a random component corresponding to the physical velocity dispersion among 
the individual stars.  Averaging over many stars in a given cluster averages
away the influence of the velocity dispersion, but has only little effect on the 
error in the radial velocity of the cluster centroid.  This quantity, discussed 
already in Paper~I (e.g.\ its Table~4), is therefore a limiting accuracy in the 
{\it average\/} astrometric radial velocity of stars in any one cluster.  For 
certain applications, effects of this noise can be lessened by averaging over 
different clusters (whose errors are not correlated), e.g., when searching for 
systematic differences between astrometric and spectroscopic radial-velocity values.

It should be noted that the selection process used to arrive at the final sample
may have a significant impact on the estimated internal velocity dispersion. The cleaning 
process successively removes those stars that deviate most from the mean cluster 
velocity, thus successively reducing $\widehat{\sigma}_v$ for the `cleaner' 
samples.  While designed to remove non-members and other outliers, this procedure
naturally affects also the mean characteristics of the remaining stars. For 
example, in the case of the Hyades, stars in the outskirts of the cluster are 
preferentially removed during the rejection procedure, meaning that the resulting 
clean samples more or less correspond to the stars within the tidal radius.  
In the case of the OB associations, we obtain velocity dispersions of about 
1~km~s$^{-1}$, a factor of two lower than the estimates for the Orion Nebula 
Cluster (Jones \& Walker \cite{jones}; Tian et al.\ \cite{tian}) and other nearby associations 
(Mathieu \cite{mathieu}). Thus, although we believe that the velocity dispersions 
reported here correctly characterize the {\em retained} samples, they are not 
necessarily representative for the cluster or association as a whole.

\subsection{Calculation of astrometric radial velocities}
\label{sec:calc}

\subsubsection{The stringent definition of `radial velocity'}
\label{sec:defin}

Recognizing the potential of astrometric radial velocities 
determined without spectroscopy, a resolution for their stringent definition was 
adopted at the General Assembly of the International Astronomical Union held in 2000.  
This resolution (Rickman \cite{rickman}) defines the geometric concept of radial velocity as 
$v_r = {\rm d}b/{\rm d}t_B$, where $b$ is the barycentric coordinate distance to 
the object and $t_{\rm B}$ the barycentric coordinate time (TCB) for light arrival 
at the solar system barycentre.  This definition is analogous to the conventional 
understanding of proper motion as the rate of change in barycentric direction 
with respect to the time of light reception at the solar-system barycentre.

In this work, we follow this IAU definition of `astrometric radial velocity'.  
The difference with respect to alternative possible definitions is on the order 
of $v_r^2/c$, with $c$ = speed of light (Lindegren et al.\ \cite{lindegren99}; Lindegren 
\& Dravins, in preparation).  Most population~I objects (including all clusters 
and associations considered in this paper) have low velocities, 
$|v_r|<50$~km~s$^{-1}$, resulting in only very small differences, $<10$~m~s$^{-1}$,
between possible alternative definitions.

\subsubsection{Radial velocities for individual stars}
\label{sec:rvind}

In the basic cluster model, the estimated radial velocity of an 
individual star is given by
\begin{equation}
    \label{eq:rv}
    \widehat{v}_{ri} = \vec{r}_i^\prime\widehat{\vec{v}}_0 \, ,
\end{equation}
where $\vec{r}_i$ is the unit vector towards star $i$ and $\widehat{\vec{v}}_0$ 
is the estimated space velocity of the cluster as a whole (actually of its 
centroid). The prime ($^\prime$) denotes the transpose of the vector. In terms 
of the equatorial coordinates $(\alpha_i,\delta_i)$ of the star we have
\begin{equation}
   \label{eq:simple}
   \widehat{v}_{ri} = \widehat{v}_{0x}\cos\delta_i\cos\alpha_i + \widehat{v}_{0y}
   \cos\delta_i\sin\alpha_i + \widehat{v}_{0z}\sin\delta_i\, ,
\end{equation}
where ($\widehat{v}_{0x}, \widehat{v}_{0y}, \widehat{v}_{0z}$) are the equatorial
velocity components as listed in Table~\ref{tab1} (or as determined by other means). 
We emphasize that Eq.~(\ref{eq:simple}) applies to {\it any\/} star that shares 
the cluster motion, irrespective of whether that star was present in the database 
used to determine the cluster motion in the first place.

The standard error $\epsilon(\widehat{v}_{ri})$ of the individual 
radial velocity is computed from
\begin{equation}
    \label{eq:rvsigma}
    \epsilon(\widehat{v}_{ri})^2 = \vec{r}_i^\prime
         {\rm Cov}(\widehat{\vec{v}}_0)\vec{r}_i + \widehat{\sigma}_v^2 \, ,
\end{equation}
where ${\rm Cov}(\widehat{\vec{v}}_0)$ is the $3\times 3$ submatrix in ${\rm 
Cov}(\widehat{\vec{\theta}})$ of all model parameters, referring to the centroid 
velocity [cf. Eq.~(A18) in Paper~II].
The first term in Eq.~(\ref{eq:rvsigma}) represents the uncertainty in the
radial component of the common cluster motion, while the second represents the
contribution due to the star's peculiar motion.

The complete covariance matrix ${\rm Cov}(\widehat{\vec{v}}_0)$ is only 
given in the electronic (extended) version of Table~\ref{tab1}. The printed
Table~\ref{tab1} gives (following the $\pm$ symbol) the standard errors 
$\epsilon(\widehat{v}_{0x})$ etc.\ of the vector components; these equal the 
square roots of the diagonal elements in ${\rm Cov}(\widehat{\vec{v}}_0)$. 
Also given for each cluster is the standard error of the radial component 
$\widehat{v}_{0r}=\vec{r}_0'\widehat{\vec{v}}_0$ of the centroid motion. 
This was computed from
\begin{equation}
    \label{eq:rv0sigma}
    \epsilon(\widehat{v}_{0r})^2 = \vec{r}_0^\prime
         {\rm Cov}(\widehat{\vec{v}}_0)\vec{r}_0 \, ,
\end{equation}
where $\vec{r}_0$ is the unit vector towards the adopted centroid position
$(\alpha_0,\delta_0)$ specified in the table. $\widehat{v}_{0r}$ can be 
regarded as an average radial velocity for the cluster as a whole, and its 
standard error (squared) can be regarded as a typical value for the first 
term in Eq.~(\ref{eq:rvsigma}). Thus, for any star not too far from the 
cluster centroid, the total standard error of its astrometric radial velocity 
can be approximately computed as 
$[\epsilon(\widehat{v}_{0r})^2+\widehat{\sigma}_v^2]^{1/2}$, using only 
quantities from the printed Table~\ref{tab1}.

\subsection{Kinematically improved parallaxes}
\label{sec:kinem}

Our maximum-likelihood estimation of the cluster space motions also produces 
estimates of the distances to all individual member stars.  A by-product of this 
moving-cluster method is therefore that individual stellar distances are 
improved, sometimes considerably, compared with the original trigonometric 
determinations.  This improvement results from a combination of the 
trigonometric parallax $\pi_{\rm trig}$ with the kinematic (secular) parallax 
$\pi_{\rm kin}=A\mu/v_t$ derived from the star's proper motion $\mu$ (scaled 
with the astronomical unit $A$) and tangential velocity $v_t$, the latter obtained 
from the estimated space velocity vector of the cluster.  The calculation of 
secular parallaxes and kinematic distances to stars in moving clusters is of 
course a classical procedure; what makes our `kinematically improved parallaxes' 
different from previous methods is that the values are derived without any 
recourse to spectroscopic data (for details, see Papers~I and II).

De Bruijne (\cite{debruijneb}) applied the present method to the Scorpius OB2
complex in order to study its HR diagram by means of the improved
distances. His `secular parallaxes' are essentially the same as our 
`kinematically improved parallaxes', being based on the same
original formulation by Dravins et al.\ (\cite{dravinsb}). The main 
differences are in the choice of rejection criteria (de Bruijne
uses $g_{\rm lim}=9$ versus our $15$) and in the practical 
implementation of the solution (downhill simplex versus our
use of analytic derivatives). De Bruijne also made extensive
Monte Carlo simulations which demonstrated that the distance 
estimates are robust against all systematic effects considered,
including cluster expansion. For the accuracy of the secular 
parallaxes, de Bruijne (\cite{debruijneb}) used a first-order formula 
(his Eqs.~16 and 17) which explicitly includes a contribution
from the (assumed) internal velocity dispersion, but neglects the 
contribution from the trigonometric parallax error (cf.\ Eq.~11
in Paper~I). By contrast, our error estimates are derived
directly from the maximum-likelihood solution (Paper~II, 
Appendix~A.3), which in principle takes into account all modelled
error sources but in practice underestimates the total error
as discussed below. As a result, our error estimates are
somewhat smaller than those given by de Bruijne (\cite{debruijneb}).

The standard errors for the estimated parallaxes 
given in this paper are the nominal ones obtained from the
maximum-likelihood estimation, which could be an underestimation of the actual errors.
Determination of realistic error estimates would require extensive Monte-Carlo
experiments based on a detailed knowledge of the actual configuration of stars,
their kinematic distributions, etc. This information is in practice unavailable
except in idealised simulations, and we therefore choose not to introduce any
ad~hoc corrections for this. As an example of the possible magnitude of
the effect, the Hyades simulations in Paper~II could be mentioned: in that 
particular case, the nominal standard errors required a correction by a factor 
1.25 to 1.28 in order to agree with the standard deviations in the actual sample.

\begin{figure}
   \begin{center}
   \resizebox{5.45cm}{!}{\includegraphics*{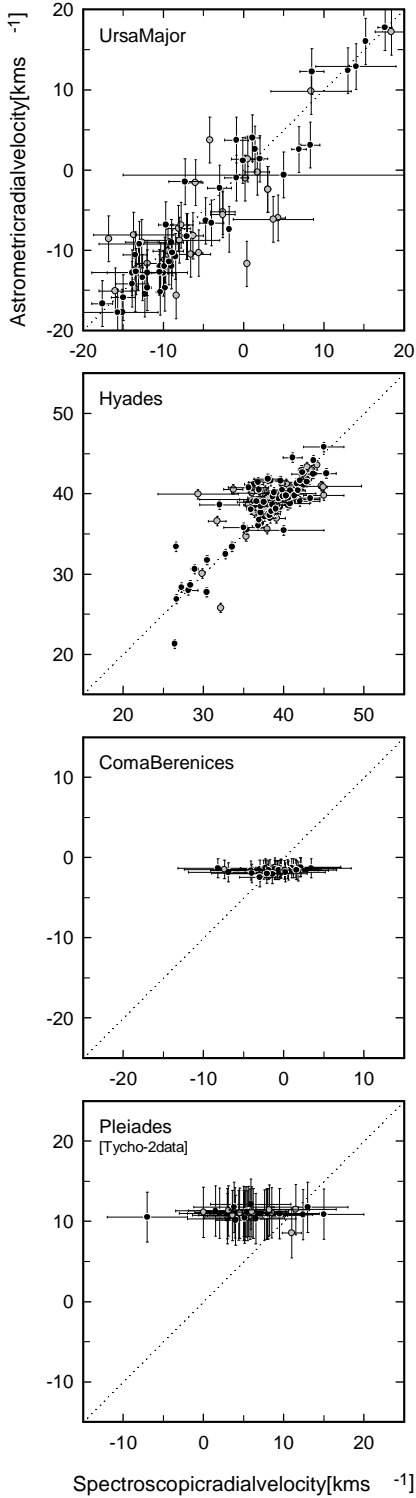}}
   \end{center}
      \caption{Astrometrically determined radial velocities compared with
spectroscopic values from the literature, for stars in four open clusters.  
The diagonal lines follow the expected relation 
$v_r({\rm astrom})\simeq v_r({\rm spectr})$.  
Black symbols denote single stars while certain or suspected binaries are in 
grey.  The top three frames are kinematic solutions obtained from Hipparcos data 
only.  In some cases, including the Pleiades (bottom), somewhat better 
accuracies are reached using data from the Tycho-2 catalogue, which incorporates 
almost a century of proper-motion data.}
   \label{fig3}
\end{figure}

\section{Radial velocities for stars in open clusters}
\label{sec:clus}

Partial data for astrometric radial velocities and kinematically improved 
parallaxes of individual stars in more nearby clusters, obtained from Hipparcos 
data, are in Table~\ref{tab2}. However, the complete listing for the more than 
1000 stars in all clusters and associations (some with also Tycho-2 solutions), 
is only in the electronic Table~\ref{tab2}.

For some clusters, Fig.~\ref{fig3} shows a comparison between {\it 
spectroscopic\/} radial velocities compiled from the literature and our 
currently determined {\it astrometric \/} values.  In all cases (also the ones 
not shown here), the astrometric values agree with the spectroscopic ones within 
the error limits, verifying the consistency of our method.  For several stars 
(especially rapidly rotating early-type ones with smeared-out spectral lines), 
the astrometric accuracies are actually substantially better than what has been 
possible to obtain from spectroscopy.  The high accuracies realized for the 
Hyades enable more detailed comparisons (Sect.~\ref{sec:comp}).

A subset of known or suspected binary stars for which Hipparcos measurements
could be perturbed are defined as stars that are either
visual binaries with magnitude difference $\Delta m < 4$~mag and a separation 
$\rho < 20$~arcsec according to HIC, the Hipparcos Input Catalogue (Turon et
al.\ \cite{turon}); known spectroscopic binaries; or flagged as 
suspected ones (identified in the Hipparcos Catalogue as a solution of type
component, acceleration, orbital, variability-induced mover, or stochastic).
In Fig.~\ref{fig3} and later, these stars are plotted in gray.

For finite astrometric accuracy, distant clusters of small angular extent 
basically yield a single astrometric velocity for all stars.  This effect is seen
in Fig.~\ref{fig3} for Coma Berenices and the Pleiades, and is similar also 
for Praesepe and the `HIP98321' clusters (not plotted).

\subsection{The Ursa Major cluster}
\label{sec:uma}

The initial sample consisted of 81 stars, being the sum of the compilations from 
Soderblom \& Mayor (\cite{soderblom}), Dravins et al.\ (\cite{dravinsb}) 
and Montes (\cite{montes}).  A few of 
the stars identified in a previous kinematic search (Dravins et al.\ \cite{dravinsb}) 
lacked spectroscopic radial velocities from the literature.  Spectroscopic 
observations of these stars by Gullberg \& Dravins (private comm.) made
with the ELODIE spectrometer at Observatoire de Haute-Provence confirmed
three of them as probable members.  The rejection procedure removed four stars, 
producing a final sample of 77.

We chose to include stars not only from the core but also from the extended halo 
(the moving group sometimes called the Sirius stream), 
in order to improve the statistical weight of the solution. Not surprisingly, 
this led to a relatively high velocity dispersion among individual stars.  Ursa 
Major may thus be viewed as a dissolved cluster moving under influence of the 
Galactic gravity field.  
Such a fate may be normal for looser open clusters reaching the age (300 Myr)
of Ursa Major (Soderblom \& Mayor \cite{soderblom}).
For convenience we use the designation `cluster' 
for the entire sample of stars.

One could imagine that moving groups like Ursa Major could be ideal targets
for the method since they have great angular extents. Unfortunately, as
Ursa Major illustrates, the high velocity dispersion
causes the errors of the estimated astrometric radial velocities to be
large. However, it should be noted that the core stars
have a much lower velocity dispersion. For instance, we get 
$\sigma_v=1.05\pm 0.22$~km~s$^{-1}$ for the 13 core (nucleus) stars 
defined by Soderblom \& Mayor (\cite{soderblom}). Wielen (\cite{wielen}) considered six core stars 
with very well determined proper motions and found the velocity dispersion 
to be on the order of 0.1~km~s$^{-1}$ (we get $0.09\pm 0.03$~km~s$^{-1}$ for 
the same six stars), adding that the much larger stream, or moving group, has 
a dispersion of $\sim 3$~km~s$^{-1}$, in good agreement with our value.
Since we assume this larger dispersion for all the stars, the
standard errors are probably overestimated for the core stars.

In the case of Ursa Major, no real gain results from the kinematically improved 
parallaxes, simply because the relative accuracy in the Hipparcos parallax values 
is already very good, given the proximity of this cluster.

The spectroscopic velocity values used for Fig.~\ref{fig3} were taken from 
Soderblom \& Mayor (\cite{soderblom}), Duflot et al.\ (\cite{duflot}), and in a few cases, also  
the ELODIE observations by Gullberg \& Dravins (private comm.). 
In the case of Duflot et al., no 
explicit error is quoted, but rather a flag which seems to correspond to a 
numerical value that is used in the Hipparcos Input Catalogue: 
those values were adopted here.

\subsection{The Hyades cluster}
\label{sec:hya}

The Hyades cluster (Melotte~25) is the classic example of a moving cluster.  Its 
kinematic distance, derived from a combination of proper motions and 
spectroscopic radial velocities, has been one of the fundamental starting points 
for the calibration of the photometric distance scale (e.g.\ Hanson \cite{hanson},
Gunn et al.\ \cite{gunn}, Schwan \cite{schwan}, and references therein). Of course, the recent 
availability of accurate trigonometric parallaxes has now superseded this method 
for distance determination.

The first detailed study of the distance, structure, membership, dynamics and 
age of the Hyades cluster, using Hipparcos data, was by Perryman et al.\ (\cite{perryman98}).  
{}From a combination of astrometric and spectroscopic radial velocity data, using 
180 stars within a radius of 20~pc, they derived the  
space velocity for the cluster centroid, 
$\vec{v}_0=(-6.32,+45.24,+5.30)$~km~s$^{-1}$.
They also estimated that the true internal 
velocity dispersion, near the centre of the cluster, is in the range 0.2 to 
0.3~{km~s$^{-1}$}. 

Our initial sample of stars was selected from the final membership assigned by 
Perryman et al.\ (\cite{perryman98}), viz.\ 197 stars classified as probable members (S~=~1 
in their Table~2): this equals our sample Hy0 in Paper~II.

In earlier kinematic studies of the Hyades, systematic errors in the proper 
motions have been of major concern, and the probable cause of discrepant 
distance estimates.  In our application, the solution is also sensitive to such 
errors, but we expect that the high internal consistency of the Hipparcos proper 
motion system, and its accurate linking to the (inertial) extragalactic 
reference frame (Kovalevsky et al.\ \cite{kovalevsky}), have effectively eliminated that 
problem.

Using the methods described in Paper~II, the astrometric radial velocity and its 
standard deviation for each individual star is obtained through 
Eqs.~(\ref{eq:rv}) and (\ref{eq:rvsigma}), even for stars not retained in the 
final sample.  Individual parallaxes follow directly from the estimation 
procedure, see Table~\ref{tab2}.

The analysis was repeated using proper motions from Tycho-2  
instead of the Hipparcos Catalogue.  Being based on observations covering a much 
longer time span, the Tycho-2 data are expected to be more precise for binaries
with periods from a few years to $\sim 100$~years. By coincidence, the procedure 
of cleaning the sample rejected the same number of stars as in the Hipparcos case, 
albeit different ones (electronic Table~\ref{tab2}). Compared to the Hipparcos solution
($0.49\pm 0.04$~km~s$^{-1}$), we get a smaller velocity dispersion using Tycho-2
($0.34\pm 0.03$~km~s$^{-1}$). It can be noted that Makarov et al.\ (\cite{makarov00})
investigated this dispersion in the Hyades as a test of the 
Tycho-2 proper motions, finding a dispersion close to our latter value.  The two solutions 
for the cluster centroid velocity are equal to within their uncertainties. 
However, in the radial direction there is a systematic difference of 
$\simeq 0.9$~km~s$^{-1}$, in the sense that the astrometric radial-velocity 
values from Tycho-2 are smaller than those from the Hipparcos Catalogue (while
the kinematically improved parallaxes from Tycho-2 place the stars at slightly 
greater distances).  We have no 
obvious explanation for this shift, which has analogues for other clusters 
when comparing Hipparcos and Tycho-2 solutions. Possibly, it reflects the 
influence of subtle systematic effects in the proper-motion data, which border 
on the measurement precision. If this is the case, greater confidence should be 
put on the solution based on the Hipparcos data, as the Tycho-2 system of
proper motions was effectively calibrated onto the Hipparcos system.  Future 
space astrometry missions should be able to clarify these matters. 

It follows from Eq.~(\ref{eq:rv}) that any possible bias in the estimated 
space velocity $\widehat{\vec{v}}_0$ has only a small influence on the 
relative astrometric radial velocities in a given cluster, if its angular
extent is not too large. Thus the astrophysical differences would still 
show up as systematic trends when the astrometric radial
velocities are compared with spectroscopic values. In such a comparison, the bias in
space velocity would mainly introduce a displacement of the zero-point,
as mentioned above.

\begin{figure}
   \resizebox{\hsize}{!}{\includegraphics*{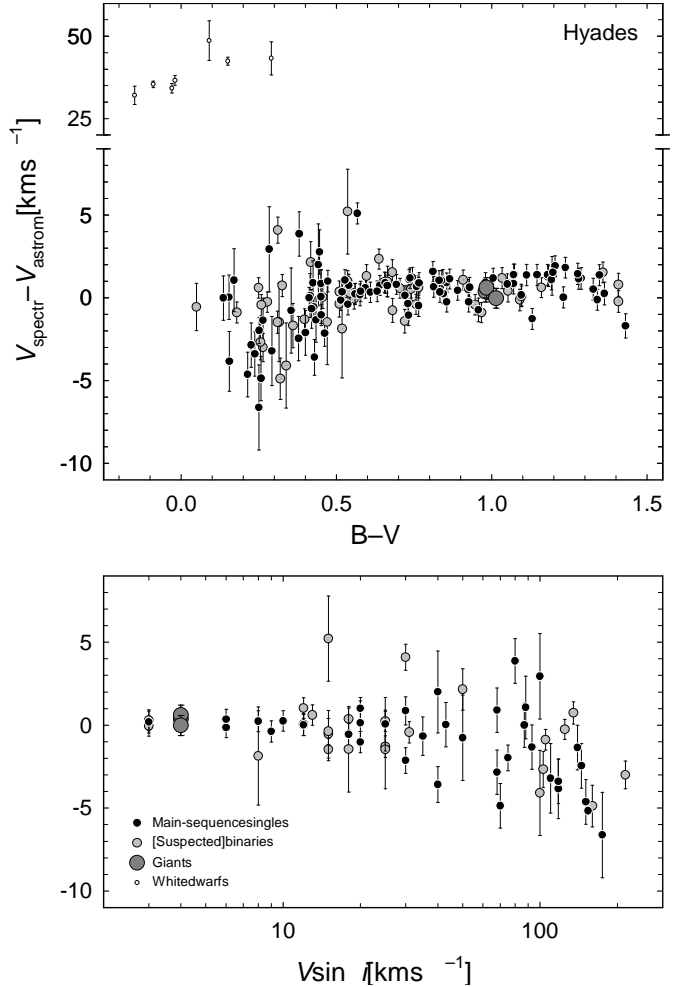}}
      \caption{The Hyades: Differences between spectroscopic radial-velocity 
values from the literature, and current astrometric determinations.  Systematic 
differences depend on spectral type, and on the projected stellar rotational 
velocity $V \sin i$. (These dependences are correlated since rapid rotation 
dominates for early-type stars.) An increased blueshift of spectral lines in 
stars somewhat hotter than the Sun ($B-V\simeq 0.3$--$0.5$) is theoretically 
expected due to their more vigorous surface convection, causing greater 
convective blueshifts.  Gravitational redshifts of white-dwarf spectra place 
them far off main-sequence stars.  The error bars show the combined 
spectroscopic and astrometric errors; stars with errors $>3$~km~s$^{-1}$ are 
omitted (except for white dwarfs), as are stars that were not retained by 
kinematic solutions from both Hipparcos and Tycho-2 data.}
   \label{fig4}
\end{figure}

\subsubsection{Hyades: Comparison with spectroscopic data}
\label{sec:comp}

The astrophysical potential of astrometric radial velocities begins to appear 
when the accuracy is sufficiently high to detect differences relative to 
spectroscopic values (e.g., Dravins et al.\ \cite{dravinsc}; Dravins \cite{dravinsa}, 
and references therein).  Such differences
are expected due to stellar surface convection (`granulation'): most photons from a 
stellar surface are emitted by hot and rising (thus locally blueshifted) 
convective elements, which contribute a greater number of photons than the cool, 
dark and sinking areas.  The resulting statistical bias causes a {\it convective 
blueshift\/}, theoretically expected to range from some 0.2~km~s$^{-1}$ in 
red dwarfs, $\simeq 0.4$~km~s$^{-1}$ in the Sun, to 1.0~km~s$^{-1}$ in 
F-type stars with their more vigorous surface convection, the precise amount 
varying among different spectral lines with dissimilar conditions of formation.  
{\it Gravitational redshifts\/} are expected to vary greatly between giants 
($<0.1$~km~s$^{-1}$), main-sequence stars (0.5--1~km~s$^{-1}$), and white dwarfs 
(perhaps 30~km~s$^{-1}$), with almost identical redshifts throughout the 
spectra.  Additional effects enter for pulsating stars, stars with expanding 
atmospheres, and such with other spectral complexities.

Our current accuracies permit such studies to be made for the Hyades, and 
perhaps marginally for a few other clusters.  In Fig.~4 we show the 
difference between astrometric radial velocities and spectroscopic measurements 
from the literature.  The latter values are taken from the compilation by 
Perryman et al.\ (\cite{perryman98}), which mostly are measurements by Griffin et al.\ 
(\cite{griffinrf}).  The values used here are their original measurements, i.e.\ {\it not\/} 
applying any zero-point or other spectral-type dependent `corrections' 
(such as were later applied in a 
convergent-point analysis for the Hyades from the same data by Gunn et al.\ 
\cite{gunn}).  The plot also includes white dwarfs, whose astrometric velocities are 
from Eq.~(\ref{eq:rv}), and where the spectroscopic data are `weighted values' 
(H$\alpha$ weighted with twice the weight of H$\beta$) from Reid (\cite{reid}).

The errors for Fig.~4 were calculated as the quadratic sum of the spectroscopic
and astrometric uncertainties, where Eq.~(\ref{eq:rvsigma}) was used for the
latter. Most of the errors are due to the spectroscopic measurements, and it can 
be noted how the scatter is greater for the [suspected] binary stars.

Over much of the main sequence, convective blueshifts and gravitational 
redshifts partly cancel one another: an increased convective blueshift in hotter 
stars is partly balanced by an increased gravitational redshift in these more 
massive stars.  Nevertheless, it is theoretically expected that the strong 
increase in the vigour of surface convection for middle F-stars ($B-V\simeq 0.4$) 
should blueshift their spectra by $\simeq 1$~km~s$^{-1}$ relative to those of 
later-type G or K-type stars ($B-V\simeq 1.0$).  For yet earlier-type stars, there 
do not yet exist any detailed theoretical models in the literature from which
the convective shift can be reliably predicted.

We believe the expected effects are visible in Fig.~4.  There is clearly a 
gradient in the relevant spectral range ($B-V$ between 0.4 and 0.7), with roughly the 
theoretically expected sign and magnitude of the effect.  The trend seems to 
continue towards even earlier types.

This is not the first time spectral-type dependent radial velocities are seen: 
a trend of increased spectral blueshift in earlier-type stars was already 
suggested from residuals in the convergent-point solution by Gunn et al.\ 
(\cite{gunn}).  A difference between the wavelength scale of giants and dwarfs, 
suggesting differences in gravitational redshift, was noted from velocity 
histograms of giants and dwarfs, respectively, in the open cluster NGC3680 by 
Nordstr{\"o}m et al.\ (\cite{nordstrom}).

Both of these works raise an important point relating to the sample
selection. Spectroscopic velocities are usually important for the 
determination of membership probabilities, which are therefore in
principle affected by systematics of the kind shown in Fig.~4. Spectral 
shifts should therefore be taken into account, lest they influence
the membership determination and hence the final result, including 
the spectral shifts themselves. Our initial Hyades sample is based on
that of Perryman et al.\ (\cite{perryman98}), who used their compilation of 
spectroscopic radial velocities to compute membership probabilities. 
Given the relatively large spectroscopic uncertainties for the early-type
stars, this effect probably did not affect the present Hyades sample.
However, as long as the mean spectral shifts remain unknown, e.g.\ as
a function of spectral type along the main sequence, it would be
necessary to downweight the more precise spectroscopic velocities in
order to avoid possible selection effects related to the spectral
shifts.

The errors in the spectroscopic velocities in several of the hottest (and often 
rapidly rotating) stars are large, making conclusions in that part of the 
diagram difficult.  For such stars with often complex spectra and perhaps 
expanding atmospheres, the concept of spectroscopic radial velocity must be 
precisely defined, if studies on the sub-km~s$^{-1}$ are to be feasible (cf.\ 
Andersen \& Nordstr{\"o}m \cite{andersen}; Griffin et al.\ \cite{griffinrem}).

\begin{figure}
   \resizebox{\hsize}{!}{\includegraphics*{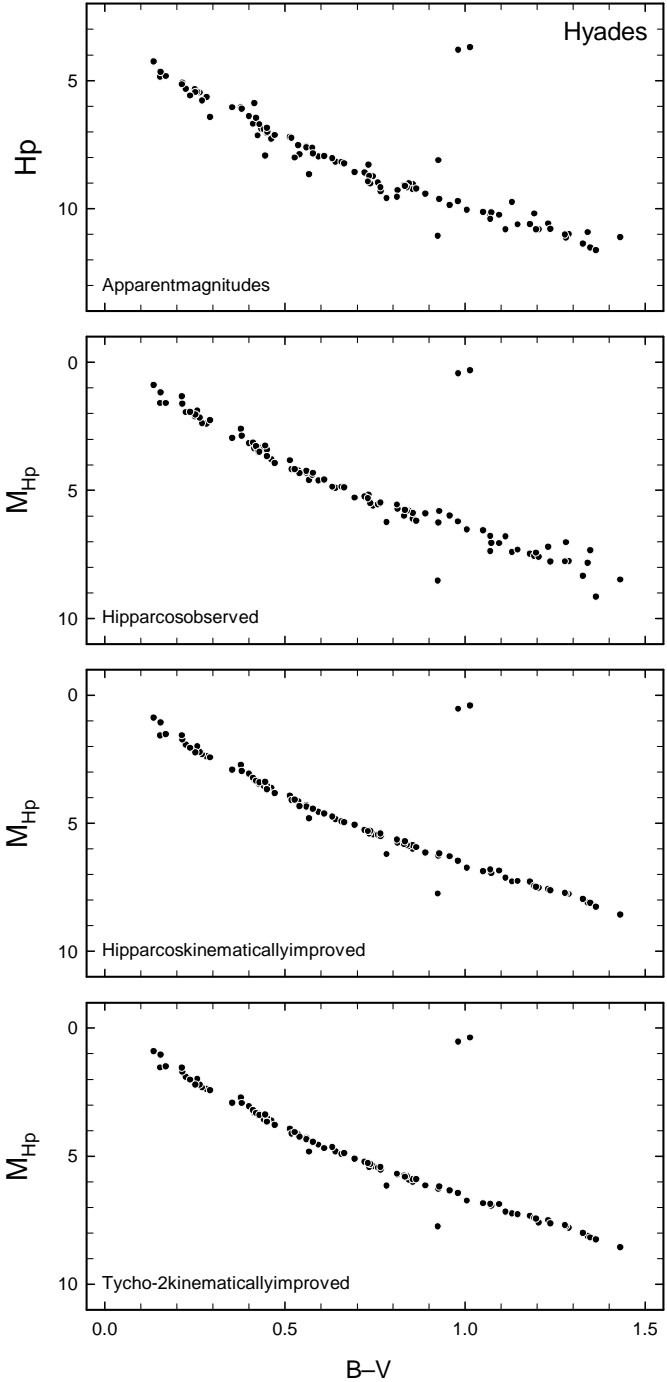}}
      \caption{The Hyades: Improved definition of the Hertzsprung-Russell 
diagram from kinematically improved parallaxes.  {}From top: (a) Apparent 
magnitudes as measured by Hipparcos; (b) Absolute magnitudes from Hipparcos 
parallaxes tighten the main sequence since the cluster depth is resolved; (c) 
Absolute magnitudes from kinematically improved parallaxes computed from 
Hipparcos data greatly improve the definition of the main sequence (especially 
for the fainter stars), further marginally improved by the use of Tycho-2 data 
(d). This permits searches for fine structure in the HR-diagram, and also 
confirms the validity of the kinematic solution for the radial-velocity 
determinations. Only single stars, retained by the kinematic solutions from both 
Hipparcos and Tycho-2 data, are plotted.}
   \label{fig5}
\end{figure}

\subsubsection{Hyades: The Hertzsprung-Russell diagram}

Already the trigonometric parallaxes from Hipparcos yield quite accurate 
absolute magnitudes, enabling a precise Hertzsprung-Russell diagram to be 
constructed.  Our kinematically improved parallaxes permit this to be carried 
further, also verifying the working of our mathematical methods.

Figure~5 shows the gradual improvements in the definition of the Hyades main 
sequence with successively better data.  The top frame shows the apparent 
magnitudes (i.e., effectively placing all stars at the same mean distance); the 
second frame shows the improvement from Hipparcos having been able to resolve 
the depth of the cluster; the third frame uses our kinematically improved 
parallaxes which, especially for the fainter stars, significantly improve the 
definition of the main sequence.  This is further marginally improved by the use 
of Tycho-2 data in the bottom frame.  The plot only shows those stars that were 
retained in both the Hipparcos and Tycho-2 solutions, and excludes [suspected] 
binary stars as defined in Sect.~\ref{sec:clus}.

Besides permitting searches for fine structure in the HR diagram, this also 
confirms the validity of the kinematic solution for the radial-velocity 
determinations: since no photometric information was used in the solution, such
an improvement could hardly be possible unless the underlying physical model
is sound.  With the kinematically improved parallaxes the error in 
$M_{\it Hp}$ is typically only $\sim 0.03$~mag, much smaller than the symbol 
size in Fig.~\ref{fig5}.

In addition to the two giants retained in the solutions, a few stars lie off the 
main sequence.  Below it is HIP~10672 at $B-V = 0.567$, a single star quite far 
(some 30~pc) from the cluster centre; HIP~17962 at $B-V = 0.782$, an eclipsing binary 
containing a hot white dwarf (Nelson \& Young \cite{nelson}) which causes a displacement 
towards the blue; and HIP~19862 at $B-V = 0.924$, with an uncertain colour index 
in the Hipparcos Catalogue ($\sigma_{B-V}=0.301$) -- the value $B-V = 1.281 \pm 0.009$
given in the Hipparcos Input Catalogue (Turon et al.\ \cite{turon}) would place it exactly on
the main sequence.  
All these stars have an uncertainty in $M_{\it Hp}$
of 0.05~mag or less, meaning that they are not misplaced vertically.
Simulations of a Hyades-type cluster by Portegies Zwart et al.\ (\cite{portegies}) showed 
that a few stars end up 
below the main sequence as a result of binary interaction leading to blueward 
displacements.  {}From Fig.~5 it is difficult to tell where the turnoff point 
really is: some stars to the far left may be blue stragglers. 

The rest of the cluster stars lie practically on a single curve, which can be considered 
a confirmation of the parallax improvement.  It is not clear whether the 
remaining spread of the main sequence in $M_{\it Hp}$ is real or can be accounted 
for by uncertainties in $B-V$, although these are small.  Effects such as 
differential reddening within the cluster seem unlikely: Taylor (\cite{taylor}) 
found only a very small colour excess $E(B-V)=0.003 \pm 0.002$~mag for the 
Hyades.

Improved absolute magnitudes were also determined by de Bruijne et al.\ (\cite{debruijned})
based on our original method (Dravins et al. \cite{dravinsb}). Lebreton
et al.\ (\cite{lebretonea}) compared our kinematically improved parallaxes to those
by de Bruijne et al., finding excellent agreement in all values, except for 
one star (HIP~28356).  We note that this particular star is the one located the 
furthest from the cluster centre, and is also one where our cleaning procedure 
removed it from the Tycho-2 solution (although it was retained in Hipparcos 
data).  It may be a long-period binary whose photocentric motion causes a 
deviation in the modulus of the measured proper motion, if not in its direction.

For further discussions of the post-Hipparcos HR diagram for the Hyades, see 
Perryman et al.\ (\cite{perryman98}), Madsen (\cite{madsen99}), 
Castellani et al.\ (\cite{castellani}), de Bruijne et 
al.\ (\cite{debruijned}), and Lebreton (\cite{lebreton}), where the latter four have used the
improved parallaxes.

\subsection{The Coma Berenices cluster}

The Coma Berenices sample is made up of the 40 Hipparcos stars in Odenkirchen et 
al.\ (\cite{odenkirchen}).  This sample includes four stars that, while slightly beyond their 
selected limit for membership, nonetheless were considered to `very probably 
also belong to the cluster'.  
Since the small number of stars made the solution unstable already after
rejecting two of them, the results in Table~\ref{tab1} are given for the full
sample ($g_{\rm lim}=\infty$).
Although the typical errors in the astrometric radial velocities are only 1.2~km~s$^{-1}$,
the precision of published spectroscopic values is generally insufficient for
meaningful comparisons. 

The kinematically improved parallaxes produce only a slight improvement in the 
HR diagram at the red end of the main sequence (Fig.~\ref{fig6}).

\subsection{The Pleiades cluster}

The sample contains 60 stars from van Leeuwen (\cite{vanleeuwen99}, and private comm.). 
The stars are too few and/or the cluster too distant for the basic cluster model to give 
a direct estimate for the velocity dispersion. It has instead been estimated 
by the procedure described in Appendix~A.4 of Paper~II.
No improvements to the parallaxes result from the kinematic 
solution, since our method is unable to resolve the depth of this cluster; 
it therefore in essence ascribes the same distance to every star.
The small angular extent means that also the astrometric radial velocity is
practically the same for all stars (Fig.~\ref{fig3}, bottom).  

\begin{figure}
   \resizebox{\hsize}{!}{\includegraphics*{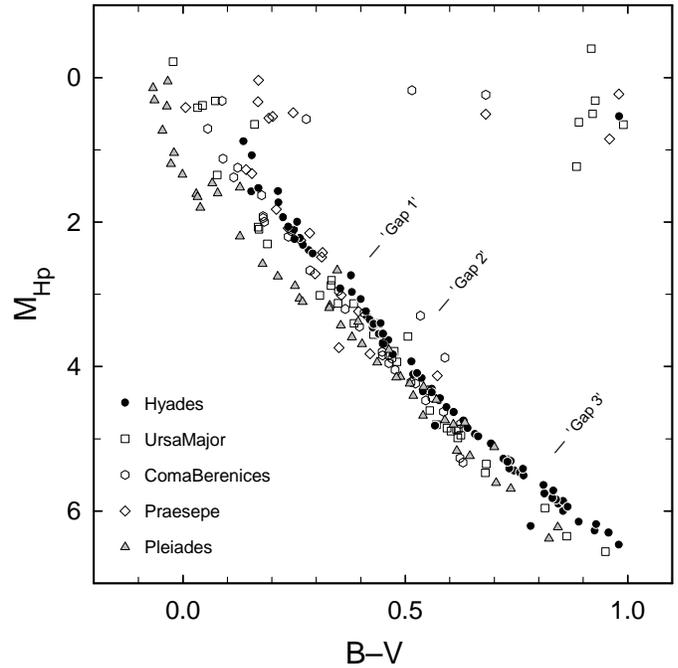}}
   \caption{Hertzsprung-Russell diagram for single stars in the better-defined 
open clusters, obtained using kinematically improved parallaxes from 
Hipparcos data.  {\it Hp} magnitudes are given since these are more precise than 
ground-based $V$ ones, and since {\it Hp} values are available for all stars in this 
sample (values for $M_{\it Hp}$ fall quite close to those of $M_V$).  The random errors in 
these kinematically improved parallaxes are lower by factors of typically 2 or 3 
compared with the original Hipparcos values, and the absolute magnitudes are 
correspondingly more precise, beginning to reveal fine structures in the 
HR diagram.  
For the Hyades, de Bruijne et al.\ (\cite{debruijnec}, \cite{debruijned}) suggested the existence of two 
underpopulated main-sequence segments around $B-V\simeq 0.38$ and $0.5$, 
identified as `B{\"o}hm-Vitense gaps', theoretically predicted due to changing 
efficiencies of stellar convection at the corresponding temperatures.  However, 
these gaps are not seen in other clusters, and their `existence' is consistent 
with small-number statistics causing random clustering.  This probably also 
applies to the apparent `Gap 3' at $B-V\simeq 0.8$.}
   \label{fig6}
\end{figure}

As seen in Fig.~\ref{fig6}, the Pleiades main sequence occupies a position at 
the lower edge of the distribution for the different clusters.
The Pleiades cluster is at the focus of an ongoing debate concerning possible
localized systematic errors in the Hipparcos parallaxes (see e.g.\ 
Pinsonneault et al.\ \cite{pinsonneault98}, \cite{pinsonneault00}; 
Robichon et al.\ \cite{robichon}; Narayanan \& Gould \cite{narayanan}; 
van Leeuwen \cite{vanleeuwen99}, \cite{vanleeuwen00}; Paper~II; Stello \& Nissen \cite{stello}). 
If such errors were present in our input data,
they would not be detected by the present maximum-likelihood method, but would
affect also the kinematically improved parallaxes. Consequently, the present 
results provide no direct new information towards the resolution of this issue.

\subsection{The Praesepe cluster}

The investigated sample was based on 24 stars from van Leeuwen (\cite{vanleeuwen99}, and 
private comm.).  As for the preceding two clusters, the solution places the 
stars at practically 
the same distance.  The resulting mean astrometric radial velocity has an error 
of $\simeq 15$~km~s$^{-1}$.  While demonstrating the applicability of the 
method, this present accuracy is insufficient for detailed stellar 
studies.

\subsection{A composite HR diagram}

Kinematically improved parallaxes from the different clusters enable a very 
detailed comparison between the main sequences of different clusters.  Such a 
Hertzsprung-Russell diagram for five nearby clusters is in Fig.~\ref{fig6}.  
We again stress that, while our kinematic solution reduces the random noise, it 
does not address any possible systematic effects and therefore cannot decide 
whether, e.g.\ the systematic shifts in luminosity between different clusters 
are caused by astrophysical or by instrumental effects.

However, the very low noise level permits to search for morphological fine 
structures in the HR diagram.  {}From post-Hipparcos data for the Hyades, de 
Bruijne et al.\ (\cite{debruijnec}, \cite{debruijned}) suggested the existence of two underpopulated 
main-sequence segments around $B-V$ approx 0.38 and 0.5, identified as 
`B\"{o}hm--Vitense gaps', 
theoretically predicted due to changing efficiencies of stellar 
convection at temperatures corresponding to those particular
colours.  However, these gaps are not seen 
in other clusters, and their `existence' is consistent with small-number 
statistics causing random clustering.  This probably also applies to the 
apparent `Gap~3' at $B-V \simeq 0.8$.  (Of course, the detectability of such 
gaps depends also on the precision in the other axis, i.e.\ the colour index).
Although some authors have suggested a possible presence of such `gaps'
(e.g., Rachford \& Canterna \cite{rachford}), extensive analyses of field stars,
using Hipparcos parallaxes, failed to show any (Newberg \& Yanny \cite{newberg}).

A certain fine structure (wiggles, etc.) is theoretically expected in the HR-
diagram (e.g., Siess et al.\ \cite{siess}); possible hints of that are becoming visible 
for the Hyades.

\begin{table*}
\caption{Data for individual stars for the better-defined clusters. Estimated 
radial velocities and their standard errors are derived from Eqs.~(\ref{eq:rv}) 
and (\ref{eq:rvsigma}), using the adopted solutions in Table~\ref{tab1} and 
the corresponding covariances. Columns: 
HIP~$=$~Hipparcos Catalogue number, $\widehat{v}_{r,\rm Hip}=$~astrometric 
radial velocity [km~s$^{-1}$] obtained from the kinematic solution using data 
from the Hipparcos main catalogue; $\widehat{\pi}_{\rm Hip}=$~estimated parallax 
[mas] from the kinematic solution using data from the Hipparcos main catalogue; 
$\epsilon(\widehat{\pi}_{\rm Hip})=$~standard error [mas] of this estimated 
parallax. Part~1: data for Ursa Major and Hyades.  The complete 
table for all clusters and associations, including results and errors obtained 
from both Hipparcos and Tycho-2 data, is available in electronic form.}
\flushleft \tiny
\begin{tabular}{*{19}{@{\extracolsep{9pt}}r}@{\extracolsep{3pt}}r} 
\hline
\noalign{\smallskip}
HIP & $\hat{v}_{r,\rm Hip}$ & $\hat{\pi}_{\rm Hip}$ & 
$\epsilon(\hat{\pi}_{\rm Hip})$\hspace{-9pt} & \hspace{16pt} &
HIP & $\hat{v}_{r,\rm Hip}$ & $\hat{\pi}_{\rm Hip}$ & 
$\epsilon(\hat{\pi}_{\rm Hip})$\hspace{-9pt} & \hspace{16pt} &
HIP & $\hat{v}_{r,\rm Hip}$ & $\hat{\pi}_{\rm Hip}$ & 
$\epsilon(\hat{\pi}_{\rm Hip})$\hspace{-9pt} & \hspace{16pt} &
HIP & $\hat{v}_{r,\rm Hip}$ & $\hat{\pi}_{\rm Hip}$ & 
$\epsilon(\hat{\pi}_{\rm Hip})$\hspace{-9pt} & \\
\noalign{\smallskip}
\hline
\noalign{\smallskip}
\noalign{\smallskip}
\noalign{\smallskip}
\noalign{\bf\footnotesize Ursa Major}
\noalign{\smallskip}
\noalign{\smallskip}
      2213   &   3.10   &   6.6    &   0.6    &&  38228  & --17.73 &   45.8  &   
0.9   && 61100  &--12.17  &    41.1  &  2.9   &&  77233   &   2.61   &   21.3   
&   0.8   \\
      8486   &   3.75   &   42.6   &   3.8    &&  42438  & --15.45 &   70.0  &   
0.7   && 61481  &--11.96  &    38.1  &  1.1   &&  80337   &  12.39   &   77.6   
&   0.9   \\
      8497   &   3.74   &   42.4   &   0.9    &&  43352  & --6.85  &   14.0  &   
1.1   && 61621  &--1.00   &    29.0  &  0.7   &&  80686   &  12.26   &   82.6   
&   0.6   \\
     10403   & --7.28   &   24.4   &   0.9    &&  46298  & --12.80 &   19.1  &   
0.9   && 61946  &--11.83  &    43.0  &  0.8   &&  80902   & --8.57   &   36.7   
&   0.6   \\
     10552   & --0.98   &   28.9   &   2.7    &&  48341  & --12.85 &   16.1  &   
0.8   && 62512  &--11.65  &    40.8  &  3.0   &&  82780   &   4.00   &   7.2    
&   0.6   \\
     17874   &   1.39   &   15.8   &   0.6    &&  48356  & --10.57 &   11.6  &   
0.8   && 62956  &--11.42  &    40.3  &  0.6   &&  83988   & --2.42   &   46.5   
&   1.8   \\
     18512   & --5.98   &   63.3   &   2.0    &&  49593  & --16.64 &   35.7  &   
0.8   && 63008  &--8.76   &    28.3  &  1.5   &&  83996   & --2.42   &   46.1   
&   2.8   \\
     19655   & --11.96  &   38.5   &   1.3    &&  49929  & --15.06 &   15.1  &   
0.8   && 63503  &--11.21  &    40.0  &  0.6   &&  87079   & --6.33   &   25.1   
&   0.6   \\
     19855   & --8.28   &   47.8   &   1.1    &&  50335  & --15.88 &   12.4  &   
0.8   && 64405  &--9.01   &    11.5  &  1.2   &&  88694   &  16.02   &   57.6   
&   0.8   \\
     19859   & --8.28   &   47.1   &   1.1    &&  51814  & --15.21 &   37.8  &   
0.6   && 64532  &--10.78  &    38.9  &  0.7   &&  91159   &   9.81   &   28.9   
&   1.5   \\
\noalign{\smallskip}
     21818   & --11.69  &   74.6   &   1.2    &&  53910  & --14.68 &   41.1  &   
0.6   && 65327  &--10.50  &    39.8  &  1.4   &&  94083   & --6.60   &   36.6   
&   0.5   \\
     23875   & --9.03   &   36.7   &   0.8    &&  53985  & --14.19 &   85.7  &   
1.4   && 65378  &--10.34  &    41.7  &  0.6   &&  96258   &   1.18   &   39.2   
&   0.5   \\
     25110   & --11.98  &   47.7   &   0.5    &&  55454  & --6.16  &   75.8  &   
1.7   && 65477  &--10.30  &    40.2  &  0.6   &&  101027  &  17.21   &   33.0   
&   0.9   \\
     27072   & --6.83   &  111.5   &   0.6    &&  56154  & --7.38  &   15.7  &   
0.8   && 66459  &--8.09   &    91.7  &  1.2   &&  103738  &  17.75   &   14.5   
&   0.8   \\
     27913   & --14.76  &  115.4   &   1.1    &&  57283  & --5.56  &   9.4   &   
0.7   && 69989  &--2.26   &    38.2  &  0.8   &&  106481  &   2.56   &   26.2   
&   0.5   \\
     28954   & --14.70  &   64.7   &   0.9    &&  57548  & --9.22  &  299.9  &   
2.2   && 71876  &--8.19   &    23.5  &  0.6   &&  110091  &  12.91   &   24.1   
&   0.9   \\
     30277   & --5.20   &   13.9   &   0.6    &&  58001  & --13.38 &   39.0  &   
0.7   && 72944  &--0.61   &   101.7  &  1.7   &&  112460  &   1.37   &  198.2   
&   2.0   \\
     30630   & --15.64  &   68.3   &   1.1    &&  59496  & --12.73 &   35.3  &   
1.2   && 73996  &--1.45   &    50.9  &  0.8   &&          &          &          
&         \\
     32349   & --10.11  &  379.2   &   1.6    &&  59514  & --12.79 &   65.6  &   
1.5   && 75312  &--1.58   &    53.8  &  1.2   &&          &          &          
&         \\
     36704   & --17.72  &   50.8   &   1.3    &&  59774  & --12.66 &   40.0  &   
0.6   && 76267  &--0.27   &    43.6  &  0.8   &&          &          &          
&         \\
\noalign{\smallskip}
\noalign{\smallskip}
\noalign{\bf\footnotesize Hyades}
\noalign{\smallskip}
\noalign{\smallskip}
     10672   &  21.30   &   17.0   &   0.3    &&  19870  &  37.88  &   20.6  &   
0.3   && 20711  & 38.60   &    21.7  &  0.3   &&  21474   &  40.54   &   20.6   
&   0.3   \\
     12709   &  25.79   &   54.0   &   0.8    &&  19877  &  38.51  &   21.5  &   
0.3   && 20712  & 38.83   &    20.9  &  0.3   &&  21482   &  38.65   &   52.0   
&   0.6   \\
     13600   &  27.73   &   15.3   &   0.3    &&  19934  &  37.80  &   19.7  &   
0.3   && 20741  & 39.50   &    22.2  &  0.4   &&  21543   &  40.69   &   19.7   
&   0.6   \\
     13806   &  26.86   &   24.5   &   0.3    &&  20019  &  38.56  &   21.1  &   
0.3   && 20745  & 39.84   &    25.1  &  0.8   &&  21589   &  40.94   &   22.3   
&   0.4   \\
     13834   &  27.97   &   30.5   &   0.3    &&  20056  &  38.46  &   21.9  &   
0.3   && 20751  & 39.93   &    23.0  &  0.5   &&  21637   &  39.67   &   23.3   
&   0.3   \\
     13976   &  28.60   &   42.8   &   0.5    &&  20082  &  38.72  &   22.4  &   
0.5   && 20762  & 39.83   &    21.3  &  0.5   &&  21654   &  40.90   &   22.8   
&   0.4   \\
     14976   &  28.32   &   25.0   &   0.3    &&  20087  &  38.03  &   18.3  &   
0.2   && 20815  & 39.71   &    21.2  &  0.3   &&  21670   &  41.17   &   20.4   
&   0.4   \\
     15300   &  30.06   &   25.5   &   0.6    &&  20130  &  38.34  &   21.9  &   
0.3   && 20826  & 39.99   &    22.3  &  0.4   &&  21683   &  40.77   &   18.3   
&   0.3   \\
     15563   &  31.72   &   32.1   &   0.4    &&  20146  &  38.66  &   21.6  &   
0.4   && 20827  & 39.82   &    20.5  &  0.4   &&  21723   &  41.08   &   22.9   
&   0.6   \\
     15720   &  30.60   &   31.0   &   0.5    &&  20205  &  38.91  &   22.1  &   
0.3   && 20842  & 38.97   &    20.2  &  0.3   &&  21741   &  39.75   &   16.6   
&   0.3   \\
\noalign{\smallskip}
     16529   &  32.47   &   23.7   &   0.3    &&  20215  &  38.84  &   24.3  &   
0.4   && 20850  & 39.89   &    21.6  &  0.4   &&  21762   &  40.81   &   21.1   
&   0.8   \\
     16548   &  33.41   &   18.3   &   0.6    &&  20219  &  39.06  &   22.3  &   
0.3   && 20873  & 39.86   &    22.0  &  0.5   &&  22044   &  41.57   &   22.7   
&   0.4   \\
     16908   &  33.39   &   21.3   &   0.3    &&  20237  &  38.56  &   22.2  &   
0.3   && 20889  & 39.39   &    21.9  &  0.3   &&  22177   &  41.71   &   22.0   
&   0.7   \\
     17766   &  35.51   &   27.3   &   0.5    &&  20261  &  39.04  &   21.0  &   
0.3   && 20890  & 39.32   &    20.9  &  0.3   &&  22203   &  41.44   &   21.2   
&   0.4   \\
     17962   &  35.44   &   21.0   &   0.3    &&  20284  &  39.17  &   20.6  &   
0.3   && 20894  & 39.78   &    22.2  &  0.4   &&  22224   &  41.21   &   22.9   
&   0.5   \\
     18018   &  34.67   &   24.3   &   0.7    &&  20349  &  38.43  &   20.2  &   
0.3   && 20899  & 39.65   &    21.6  &  0.3   &&  22253   &  40.40   &   18.3   
&   0.4   \\
     18170   &  35.77   &   23.7   &   0.3    &&  20350  &  38.80  &   21.4  &   
0.3   && 20901  & 40.02   &    21.3  &  0.3   &&  22265   &  41.21   &   20.0   
&   0.4   \\
     18322   &  36.32   &   21.5   &   0.4    &&  20357  &  39.20  &   20.4  &   
0.3   && 20916  & 39.79   &    18.7  &  0.5   &&  22271   &  39.76   &   27.3   
&   0.5   \\
     18327   &  36.03   &   24.4   &   0.4    &&  20400  &  39.27  &   22.3  &   
0.3   && 20935  & 39.67   &    21.8  &  0.3   &&  22350   &  40.86   &   20.6   
&   0.4   \\
     18658   &  36.94   &   23.5   &   0.5    &&  20419  &  39.46  &   22.1  &   
0.5   && 20948  & 39.65   &    21.8  &  0.3   &&  22380   &  41.27   &   20.9   
&   0.4   \\
\noalign{\smallskip}
     18735   &  36.58   &   22.0   &   0.3    &&  20440  &  39.27  &   21.7  &   
0.5   && 20949  & 38.12   &    17.3  &  0.3   &&  22394   &  40.18   &   20.4   
&   0.4   \\
     18946   &  36.76   &   21.0   &   0.4    &&  20455  &  39.04  &   21.1  &   
0.3   && 20951  & 39.65   &    22.3  &  0.4   &&  22422   &  41.65   &   20.8   
&   0.4   \\
     19082   &  36.96   &   20.8   &   0.5    &&  20480  &  38.56  &   19.8  &   
0.3   && 20978  & 39.83   &    22.0  &  0.4   &&  22505   &  41.81   &   21.9   
&   0.4   \\
     19098   &  37.16   &   22.1   &   0.4    &&  20482  &  38.82  &   19.2  &   
0.3   && 20995  & 39.95   &    22.3  &  0.4   &&  22524   &  41.72   &   20.3   
&   0.4   \\
     19148   &  37.44   &   20.9   &   0.3    &&  20484  &  39.17  &   20.8  &   
0.3   && 21008  & 39.46   &    19.1  &  0.3   &&  22550   &  42.15   &   21.3   
&   0.4   \\
     19207   &  37.56   &   21.6   &   0.4    &&  20485  &  39.27  &   24.9  &   
0.5   && 21029  & 39.93   &    21.8  &  0.3   &&  22565   &  41.43   &   19.3   
&   0.3   \\
     19261   &  37.65   &   22.6   &   0.3    &&  20491  &  38.04  &   18.7  &   
0.3   && 21036  & 40.15   &    22.4  &  0.3   &&  22566   &  41.85   &   16.4   
&   0.3   \\
     19263   &  37.53   &   21.7   &   0.4    &&  20492  &  39.38  &   21.0  &   
0.4   && 21039  & 39.99   &    21.7  &  0.4   &&  22654   &  41.48   &   19.2   
&   0.5   \\
     19316   &  37.91   &   20.8   &   0.5    &&  20527  &  39.46  &   22.6  &   
0.6   && 21066  & 40.34   &    21.9  &  0.4   &&  22850   &  41.61   &   15.9   
&   0.3   \\
     19365   &  35.57   &   14.7   &   0.2    &&  20542  &  39.17  &   22.1  &   
0.3   && 21099  & 39.51   &    21.7  &  0.4   &&  23069   &  42.48   &   18.0   
&   0.4   \\
\noalign{\smallskip}
     19441   &  38.15   &   28.3   &   0.5    &&  20557  &  38.59  &   23.6  &   
0.4   && 21112  & 40.22   &    19.6  &  0.3   &&  23214   &  42.44   &   23.3   
&   0.4   \\
     19504   &  37.66   &   22.5   &   0.3    &&  20563  &  39.12  &   22.3  &   
0.5   && 21123  & 39.87   &    22.1  &  0.4   &&  23312   &  42.94   &   19.3   
&   0.5   \\
     19554   &  38.28   &   26.5   &   0.4    &&  20567  &  39.24  &   19.8  &   
0.4   && 21137  & 40.09   &    22.8  &  0.3   &&  23497   &  41.84   &   19.1   
&   0.3   \\
     19591   &  37.03   &   25.2   &   0.4    &&  20577  &  39.27  &   21.8  &   
0.4   && 21138  & 40.13   &    21.1  &  1.0   &&  23498   &  42.90   &   18.4   
&   0.5   \\
     19781   &  38.43   &   19.8   &   0.3    &&  20605  &  39.40  &   20.8  &   
1.5   && 21152  & 40.46   &    23.8  &  0.4   &&  23701   &  43.34   &   18.5   
&   0.7   \\
     19786   &  38.57   &   21.6   &   0.4    &&  20614  &  39.06  &   21.6  &   
0.3   && 21179  & 40.37   &    21.9  &  0.6   &&  23750   &  42.67   &   18.8   
&   0.4   \\
     19789   &  37.50   &   17.6   &   0.2    &&  20635  &  38.60  &   21.1  &   
0.3   && 21256  & 39.56   &    23.2  &  0.4   &&  23983   &  43.56   &   19.0   
&   0.4   \\
     19793   &  37.31   &   22.4   &   0.3    &&  20641  &  38.62  &   22.5  &   
0.3   && 21261  & 39.89   &    21.6  &  0.5   &&  24019   &  40.81   &   17.9   
&   0.3   \\
     19796   &  38.65   &   22.0   &   0.3    &&  20648  &  39.25  &   21.8  &   
0.3   && 21267  & 40.49   &    21.8  &  0.4   &&  24116   &  42.53   &   12.4   
&   0.3   \\
     19808   &  38.58   &   21.9   &   0.5    &&  20661  &  39.48  &   21.3  &   
0.3   && 21273  & 40.37   &    22.4  &  0.5   &&  24923   &  44.17   &   17.7   
&   0.5   \\
\noalign{\smallskip}
     19834   &  38.53   &   21.3   &   0.8    &&  20679  &  39.27  &   22.9  &   
0.5   && 21317  & 40.38   &    22.0  &  0.4   &&  26382   &  44.49   &   19.7   
&   0.5   \\
     19862   &  38.46   &   21.8   &   0.5    &&  20686  &  39.17  &   22.1  &   
0.4   && 21459  & 39.43   &    23.6  &  0.3   &&  28356   &  45.83   &   13.4   
&   0.6   \\
\noalign{\smallskip}
\hline 
\end{tabular}
\label{tab2}
\end{table*}

\begin{table*}\addtocounter{table}{-1}
\caption{{\bf (Continued)} Data for individual stars for the better-defined 
clusters. Part~2: data for Coma Berenices, Pleiades and Praesepe.  The complete 
table for all clusters and associations, including results and errors obtained 
from both Hipparcos and Tycho-2 data, is available in electronic form.}
\flushleft \tiny
\begin{tabular}{*{19}{@{\extracolsep{9pt}}r}@{\extracolsep{3pt}}r} 
\hline
\noalign{\smallskip}
HIP & $\hat{v}_{r,\rm Hip}$ & $\hat{\pi}_{\rm Hip}$ & 
$\epsilon(\hat{\pi}_{\rm Hip})$\hspace{-9pt} & \hspace{16pt} &
HIP & $\hat{v}_{r,\rm Hip}$ & $\hat{\pi}_{\rm Hip}$ & 
$\epsilon(\hat{\pi}_{\rm Hip})$\hspace{-9pt} & \hspace{16pt} &
HIP & $\hat{v}_{r,\rm Hip}$ & $\hat{\pi}_{\rm Hip}$ & 
$\epsilon(\hat{\pi}_{\rm Hip})$\hspace{-9pt} & \hspace{16pt} &
HIP & $\hat{v}_{r,\rm Hip}$ & $\hat{\pi}_{\rm Hip}$ & 
$\epsilon(\hat{\pi}_{\rm Hip})$\hspace{-9pt} & \\
\noalign{\smallskip}
\hline
\noalign{\smallskip}
\noalign{\smallskip}
\noalign{\smallskip}
\noalign{\bf\footnotesize Coma Berenices}
\noalign{\smallskip}
\noalign{\smallskip}
59364   &  --1.35   &    11.0   &    0.6   &&   60123   &  --1.59   &    11.5   
&    0.6   &&   60525   &  --1.48   &    10.7   &    0.6   &&   61205   &  
--2.25   &    13.2   &    1.0    \\
59399   &  --1.27   &    10.5   &    0.8   &&   60206   &  --1.43   &    11.3   
&    0.7   &&   60582   &  --1.49   &    10.3   &    0.7   &&   61295   &  
--1.56   &    10.4   &    0.5    \\
59527   &  --1.39   &    11.4   &    0.6   &&   60266   &  --1.39   &    11.0   
&    0.6   &&   60611   &  --1.34   &    11.4   &    0.7   &&   61402   &  
--1.79   &    11.0   &    0.7    \\
59833   &  --1.34   &    10.4   &    0.6   &&   60293   &  --1.50   &    10.9   
&    0.8   &&   60649   &  --1.59   &    12.0   &    0.6   &&   62384   &  
--1.70   &    11.9   &    0.7    \\
59957   &  --1.36   &    11.3   &    0.6   &&   60304   &  --1.55   &    11.1   
&    0.7   &&   60697   &  --1.63   &    13.1   &    0.5   &&   62763   &  
--2.10   &    10.9   &    0.6    \\
60014   &  --1.85   &    12.3   &    0.6   &&   60347   &  --1.23   &    11.2   
&    0.6   &&   60746   &  --1.61   &    11.6   &    0.5   &&   62805   &  
--1.98   &    12.5   &    0.8    \\
60025   &  --1.21   &    12.2   &    0.7   &&   60351   &  --1.46   &    11.4   
&    0.5   &&   60797   &  --1.56   &    10.9   &    0.6   &&   63493   &  
--2.02   &    11.7   &    0.7    \\
60063   &  --1.33   &    11.5   &    0.7   &&   60406   &  --1.48   &    10.6   
&    0.7   &&   61071   &  --1.53   &    11.2   &    0.5   &&   64235   &  
--2.78   &    12.8   &    1.0    \\
60066   &  --1.41   &    12.3   &    0.5   &&   60458   &  --1.56   &    11.6   
&    0.6   &&   61074   &  --1.74   &    11.9   &    0.6   &&   65466   &  
--2.48   &    11.7   &    0.6    \\
60087   &  --1.21   &    11.6   &    0.5   &&   60490   &  --1.49   &    11.8   
&    0.5   &&   61147   &  --1.86   &    11.6   &    0.6   &&   65508   &  
--2.60   &    11.2   &    0.8    \\
\noalign{\smallskip}
\noalign{\smallskip}
\noalign{\bf\footnotesize Pleiades}
\noalign{\smallskip}
\noalign{\smallskip}
16217   &    9.06   &    8.6    &    0.2   &&   17289   &   11.20   &    7.9    
&    0.3   &&   17573   &   10.70   &    8.5    &    0.2   &&   17847   &   
10.98   &    8.3    &    0.2    \\
16407   &    9.63   &    8.7    &    0.2   &&   17317   &   11.40   &    8.4    
&    0.3   &&   17579   &   10.62   &    8.4    &    0.2   &&   17851   &   
10.95   &    8.7    &    0.2    \\
16635   &    9.93   &    8.2    &    0.4   &&   17325   &   12.39   &    8.7    
&    0.2   &&   17583   &   10.25   &    8.5    &    0.2   &&   17862   &   
10.85   &    8.3    &    0.2    \\
16639   &   10.17   &    8.2    &    0.3   &&   17401   &   10.92   &    8.5    
&    0.2   &&   17588   &   10.64   &    8.3    &    0.2   &&   17892   &   
11.67   &    8.4    &    0.2    \\
16753   &   10.90   &    8.7    &    0.3   &&   17481   &   12.23   &    8.3    
&    0.2   &&   17607   &   12.25   &    7.9    &    0.4   &&   17900   &   
11.16   &    8.2    &    0.2    \\
16979   &   11.18   &    8.1    &    0.3   &&   17489   &   10.69   &    8.3    
&    0.2   &&   17608   &   10.91   &    8.1    &    0.2   &&   17923   &   
11.11   &    8.2    &    0.3    \\
17000   &   11.12   &    8.4    &    0.2   &&   17497   &   11.14   &    8.4    
&    0.2   &&   17625   &   10.07   &    8.4    &    0.2   &&   17999   &   
11.10   &    8.4    &    0.2    \\
17020   &   10.30   &    8.1    &    0.3   &&   17499   &   10.77   &    8.5    
&    0.2   &&   17664   &   10.68   &    8.6    &    0.2   &&   18050   &   
10.87   &    8.5    &    0.2    \\
17034   &   10.25   &    8.6    &    0.2   &&   17511   &   11.69   &    8.1    
&    0.3   &&   17692   &   11.01   &    8.3    &    0.2   &&   18154   &   
10.84   &    8.3    &    0.3    \\
17044   &   10.36   &    8.4    &    0.3   &&   17525   &    9.81   &    8.9    
&    0.5   &&   17694   &   11.40   &    8.5    &    0.3   &&   18263   &   
11.07   &    9.1    &    0.4    \\
\noalign{\smallskip}
17091   &   10.90   &    8.8    &    0.3   &&   17527   &   10.46   &    8.6    
&    0.2   &&   17702   &   10.89   &    8.1    &    0.2   &&   18266   &   
12.38   &    8.3    &    0.4    \\
17125   &    8.93   &    8.1    &    0.3   &&   17531   &   10.63   &    7.9    
&    0.2   &&   17704   &   10.81   &    8.3    &    0.2   &&   18431   &   
11.69   &    8.6    &    0.2    \\
17168   &    8.49   &    7.5    &    0.4   &&   17547   &    8.75   &    8.9    
&    0.3   &&   17729   &   10.33   &    8.6    &    0.2   &&   18544   &   
12.83   &    9.5    &    0.3    \\
17225   &   10.89   &    8.6    &    0.3   &&   17552   &   12.05   &    9.0    
&    0.2   &&   17776   &   11.23   &    8.7    &    0.2   &&   18559   &   
11.35   &    8.2    &    0.2    \\
17245   &    9.95   &    8.0    &    0.3   &&   17572   &   11.24   &    8.5    
&    0.2   &&   17791   &   10.82   &    8.0    &    0.2   &&   18955   &   
12.06   &    8.5    &    0.3    \\
\noalign{\smallskip}
\noalign{\smallskip}
\noalign{\bf\footnotesize Praesepe}
\noalign{\smallskip}
\noalign{\smallskip}
41788   &   37.62   &    5.9    &    0.3   &&   42327   &   36.63   &    5.7    
&    0.3   &&   42556   &   36.25   &    5.5    &    0.2   &&   42952   &   
35.76   &    5.3    &    0.3    \\
42133   &   36.86   &    5.7    &    0.3   &&   42485   &   36.30   &    5.7    
&    0.3   &&   42578   &   36.19   &    5.3    &    0.2   &&   42966   &   
35.47   &    6.0    &    0.3    \\
42164   &   36.78   &    5.7    &    0.3   &&   42516   &   36.21   &    5.4    
&    0.2   &&   42600   &   36.18   &    5.0    &    0.2   &&   42974   &   
35.42   &    5.8    &    0.3    \\
42201   &   36.60   &    5.6    &    0.3   &&   42518   &   35.99   &    5.8    
&    0.3   &&   42673   &   36.02   &    5.7    &    0.3   &&   43050   &   
35.50   &    5.8    &    0.3    \\
42247   &   36.82   &    5.6    &    0.3   &&   42523   &   36.20   &    5.7    
&    0.3   &&   42705   &   36.15   &    5.9    &    0.3   &&   43086   &   
35.63   &    5.6    &    0.3    \\
42319   &   36.59   &    5.7    &    0.3   &&   42549   &   36.24   &    5.7    
&    0.3   &&   42766   &   35.70   &    5.7    &    0.3   &&   43199   &   
35.45   &    5.2    &    0.4    \\
\noalign{\smallskip}
\hline
\end{tabular}
\end{table*}

\section{Application to OB associations}
\label{sec:assoc}

The procedures of determining astrometric radial velocities were applied also to 
a number of nearby associations of young stars.  The situation is here somewhat 
different from that of the previously discussed older clusters because (at least 
some of) these younger associations may be undergoing significant expansion, or 
have otherwise complex patterns of stellar motion on levels comparable to our 
desired accuracies.  We recall that the present moving-cluster method is based 
upon measuring the rate of angular expansion or contraction: it cannot 
therefore segregate 
whether a change in angular scale occurs because the cluster is approaching or 
expanding.  While -- on the accuracy levels aimed at -- this should not be a 
problem for the older clusters, the likely expansion of young associations may 
introduce significant biases in the solution.  Another complication is that, 
since some of the associations cover large areas of sky, there is an increased 
risk for contamination of the samples by field stars.  Further, spectroscopic 
radial velocities often cannot be used to decide membership, both because they 
do not exist in significant numbers, and because their actual measurement is 
difficult for the often rapidly rotating O and B-type stars that make up much of 
these associations. For such reasons, the associations are here being treated
separately.

The predictions in Paper~I indicate that the accuracies of Hipparcos should 
enable astrophysically interesting results to be obtained for perhaps half a 
dozen of the nearer associations.  Among these, Lower Centaurus Crux, 
Upper Centaurus Lupus and Upper Scorpius form part of the larger Scorpius OB2 
complex, while the $\alpha$~Persei and `HIP~98321' associations are independent 
entities.

Except for `HIP 98321' (Sect.~\ref{sec:hip98321}) the selection of members in 
the different associations is based on data from de Zeeuw et al.\ (\cite{dezeeuw}).
In their sample selection they combine one method using Hipparcos 
positions and proper motions (de Bruijne \cite{debruijnea}), and another using Hipparcos 
positions, proper motions and parallaxes (Hoogerwerf \& Aguilar \cite{hoogerwerf}).  Although 
this could cause some contamination by outliers, simulations showed that only 
20\% of the stars in the first method are expected to be field stars, and only 
4\% in the second.  Although, in principle, our procedure for rejecting outliers 
does reduce this contamination, actually only few stars were rejected.

De Bruijne (\cite{debruijneb}) used an implementation of our original method (Dravins 
et al.\ \cite{dravinsb}) to obtain kinematically improved parallaxes for the three 
OB associations in the Scorpius OB2 complex (cf.\ Sect.~\ref{sec:kinem}). 
While the depth of the associations is not fully resolved by the Hipparcos 
parallaxes, the kinematically improved parallaxes reveal some internal 
structure. We refer to de Bruijne's work concerning the three-dimensional 
structure of the complex, although his distance estimates are slightly 
different from ours (mainly because his selection criterion, $g_{\rm lim}=9$, 
differs from our $g_{\rm lim}=15$). Based on the kinematically improved 
parallaxes, we presented the Hertzsprung-Russell diagrams of Upper Centaurus 
Lupus and Lower Centaurus Crux in Madsen et al. (\cite{madsen00}). For additional 
discussion of the HR diagrams of the complex we again refer to de Bruijne (\cite{debruijneb}).

The solutions for the associations as a whole were given in Table~\ref{tab1} 
(and its electronic version), while the results for the
individual stars are given in the electronic version of Table~\ref{tab2}.

\subsection{The Lower Centaurus Crux association}
\label{sec:lcc}

The cleaned sample consists of 179 stars with an estimated internal dispersion
of 1.1~km~s$^{-1}$. Combined with the rather small uncertainty of the cluster 
velocity, the resulting standard error for the astrometric radial velocities 
is 1.2 to 1.3~km~s$^{-1}$.  In the corresponding HR diagram, the main sequence 
becomes somewhat better defined, most noticeably in the A-star regime 
(Madsen et al.\ \cite{madsen00}), although there still remains a significant spread.

\subsection{The Upper Centaurus Lupus association}
\label{sec:ucl}

For Upper Centaurus Lupus, the rejection procedure produces a clean sample with 
218 stars with an estimated internal dispersion of 1.2~km~s$^{-1}$. The resulting
standard error of the astrometric radial velocities is 1.3~km s$^{-1}$.

The HR diagram clearly shows an improvement across the whole spectral range of 
the main sequence (Madsen et al.\ \cite{madsen00}).  
Probably, the remaining 
spread is caused by non-detected binaries, some non-members, and pre-main sequence 
objects moving onto the main sequence.  Differential reddening across the 
association and perhaps also in depth could also cause a spread of the main 
sequence, although Upper Centaurus Lupus is not believed to be as much affected 
as the other two associations in the Scorpius OB2 complex 
(de Zeeuw et al.\ \cite{dezeeuw}).

\subsection{The Upper Scorpius association}
\label{sec:usc}

The maximum-likelihood solution for the Upper Scorpius association became unstable 
after rejection of eight stars, at which point the criterion $g_{\rm max}\le 15$ 
was still not met.  We therefore choose to give results for the solution using all 
120 stars in the original sample.  The internal dispersion is in line with that of 
the previous two associations, but the larger uncertainty in the cluster velocity 
gives a higher standard error of about 1.9~km~s$^{-1}$ for the astrometric radial 
velocities.

It is difficult to judge whether the main sequence is actually better delineated 
by the kinematically improved parallaxes.  Upper Scorpius appears to be close to 
the limit of our method, due to its larger distance, smaller angular size and a 
smaller number of member stars, compared with Lower Centaurus Crux and 
Upper Centaurus Lupus.

\subsection{The Scorpius~OB2 complex}
\label{sec:sco}

Lower Centaurus Crux, Upper Centaurus Lupus and Upper Scorpius are all 
part of a larger OB complex, known as Scorpius~OB2, with similar space 
velocity vectors (Blaauw \cite{blaauw}).  Therefore, an attempt was also made to 
combine the three associations in a single solution assuming a common 
velocity vector. The resulting vector and internal dispersion are  
in Table~\ref{tab1}, but we give no results for individual stars. 

The HR diagram is visibly improved, indicating that Sco~OB2 could meaningfully 
be regarded as one single structure.  However, a combination of the 
HR diagrams from the three separate solutions is even slightly better 
defined, suggesting that Sco~OB2 is, after all, better considered as three 
separate structures.

When treating Sco~OB2 as one complex, the estimated internal velocity dispersion
is only slightly larger than for the separate solutions, and the formal 
uncertainty of the space velocity vector is remarkably small. Nevertheless,
when comparing the resulting astrometric radial velocities with those from the
previous solutions we find noticeable differences. For Lower Centaurus 
Crux (LCC) we find
$\langle v_{ri}({\rm LCC})-v_{ri}({\rm Sco~OB2})\rangle\simeq -2$~km~s$^{-1}$, 
while for Upper Centaurus Lupus and Upper Scorpius the corresponding mean 
differences are $+4$~km~s$^{-1}$ and $+10$~km~s$^{-1}$,
respectively. Such a progression of systematic differences could be expected 
if Sco~OB2 is not really one uniform complex, or if there is some internal
velocity field.  At any rate, the comparison shows that one has to be careful 
when interpreting the results for young associations: although we get 
a stable solution with small residuals when considering the whole complex,
the resulting velocities are not trustworthy.  

Thus both the HR diagrams and the radial-velocity solutions indicate that 
the Sco~OB2 complex has some internal kinematic structure that ultimately 
will need to be modelled, although it is only marginally discernible in the
present data. In Sect.~\ref{sec:expand} we discuss the possible expansion
of the associations.

\subsection{The $\alpha$~Persei association (Per~OB3)}
\label{sec:alphaper}

This $\alpha$~Per association is sometimes denoted an open cluster.  {}From 
our sample we obtain a mean astrometric radial velocity of 
$4.5 \pm 2.2$~km~s$^{-1}$.  A rather modest internal 
velocity dispersion $\sigma_v\simeq$ 0.7~km~s$^{-1}$ was found using the
procedure of Appendix~A.4 in Paper~II. The value is smaller than for the
other OB associations, and indicates that it may be reasonable to look upon 
the structure as a young open cluster instead. The velocity dispersion, 
together with the uncertainty in the solution for the cluster velocity, 
combine to give a standard error of about 2.3~km~s$^{-1}$ in the astrometric 
radial velocities of the individual stars. The parallax improvement is not 
good enough to have a visible impact on the HR diagram.

Our radial-velocity result is close to the spectroscopic values of
$\simeq +2$~km~s$^{-1}$ (Prosser \cite{prosser}), while somewhat larger than the 
$-0.9$~km~s$^{-1}$ derived from the convergence-point solution by Eggen (\cite{eggen98}).

\subsection{The `HIP~98321' association}
\label{sec:hip98321}

This possible association was recently discovered in the Cepheus-Cygnus-Lyra-Vulpecula 
region by Platais et al.\ (\cite{platais}), during a search for new star clusters from 
Hipparcos data.  They named it after the central star HIP~98321, and found 
59 probable members.  Because of the Hipparcos limiting magnitude, only O, B, 
and A-type stars are utilizable.  It was a bit surprising to find that this 
association gives a good kinematic solution despite its great distance of 307~pc; 
the reason is probably its large mean radius on the sky of $\sim 12$~degrees 
(Fig.~1).

The mean astrometric radial velocity is $-19.3 \pm 1.6$~km~s$^{-1}$ for the 
sample of all 59 stars.  Together with the estimated internal dispersion of 
2.6~km~s$^{-1}$, the standard error of the individual astrometric radial
velocities is around 3.2~km~s$^{-1}$.
These values are consistent with the somewhat uncertain spectroscopic
velocities for these early-type stars. Published values spread around 
$-15$~km~s$^{-1}$, suggesting a possible expansion of the association
compatible with its isochrone age (Table~4 in Paper~I and next section).

During the cleaning process, the maximum $g$ value was always below 
$g_{\rm lim}=15$; thus no star was removed from the original sample. 
Some contamination by outliers may nevertheless be expected due to the 
lack of spectroscopic information in the selection of the stars.
It would have been a nice confirmation of the existence of this new 
association if the improved parallaxes had given a better-defined main 
sequence, but unfortunately the improvement is not sufficient to have any 
visible effect in the HR diagram.

\begin{figure}
   \resizebox{\hsize}{!}{\includegraphics*{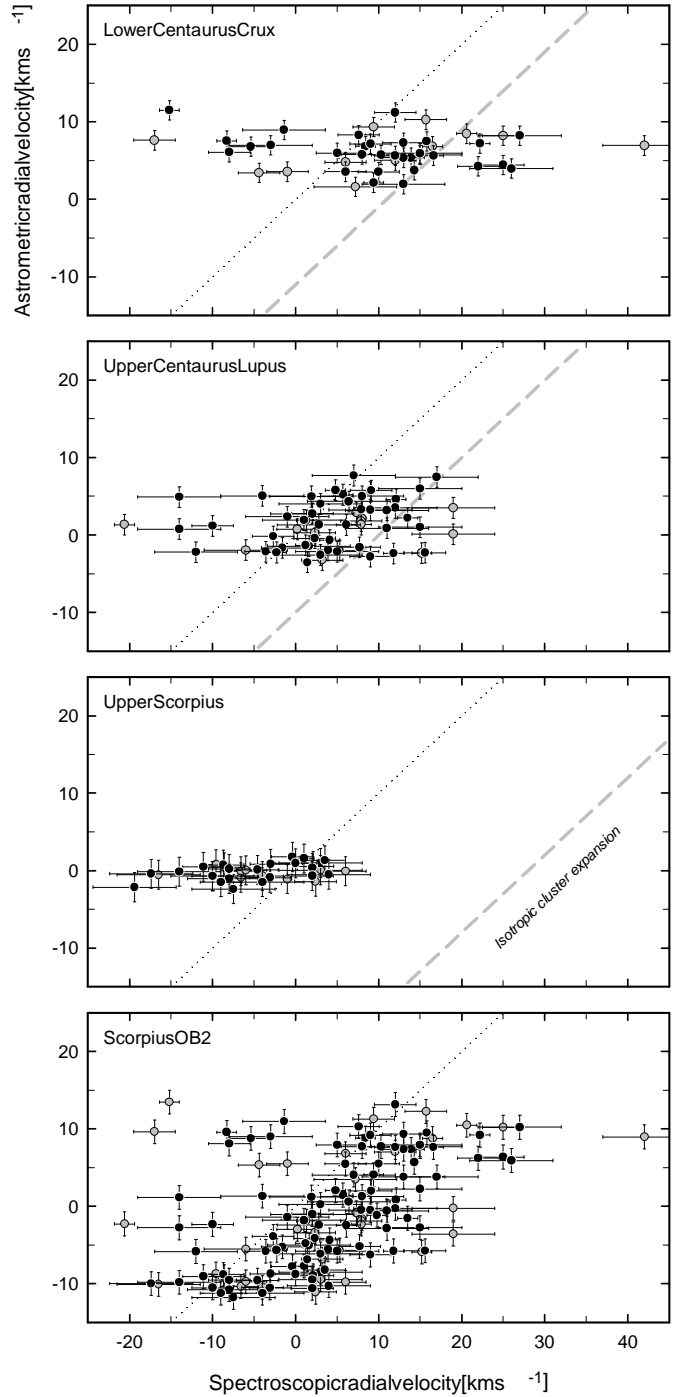}}
   \caption{Astrometric versus spectroscopic radial velocities for stars in the 
Scorpius-Centaurus group of young associations, expected to undergo kinematic 
expansion.  The top three frames show separate solutions for each subgroup.  
Assuming a rate of isotropic expansion equal to the inverse age of the cluster, 
a bias in the astrometric radial velocity would result, marked by dashed gray 
lines (the cluster's increasing angular size would be interpreted as approaching 
motion; Paper~I).  The assumed ages are 11, 14 and 5~Myr, respectively 
(de Geus et al.\ \cite{degeus}).
The bottom frame shows the solution for all 510 stars in the 
groups, treated as one entity (only stars with known spectroscopic velocities 
are plotted).  While these data do indicate \textit{some} expansion of this 
complex of young associations, the expansion of its individual parts is 
significantly slower than the naively expected rate.}
   \label{fig7}
\end{figure}

\subsection{Expanding associations?}
\label{sec:expand}

Figure~\ref{fig7} shows the astrometric versus spectroscopic radial velocities 
for stars in the Scorpius-Centaurus group of young associations, both for each 
individual subgroup, and for the complex treated as a whole. The spectroscopic 
values are those compiled in the Hipparcos Input Catalogue HIC 
(Turon et al.\ \cite{turon}). Because we are dealing with young and rapidly rotating 
early-type stars,  
the spectroscopic errors are quite large; some contamination is also expected due 
to outliers and binaries.

The astrometric radial velocities in OB associations are expected to show a 
significant bias due to expansion effects (Paper~I).  Assuming the inverse age 
of an association to be the upper limit on the relative expansion rate, the
resulting maximum bias in the astrometric radial velocity can be computed from 
Eq.~(10) in Paper~I. This effect is directly proportional to the 
distance to the stars in the association and inversely proportional to its age. 
The expansion
causes a positive shift in $v_r({\rm spectr})-\widehat{v}_{r}({\rm astrom})$: the 
cluster's increasing angular size is wrongly interpreted as approaching motion.
This [upper limit of the] expansion bias is plotted in Fig.~\ref{fig7} together 
with the spectroscopic and astrometric velocities. We have not been able
to observe any correlation between distance and expansion with the present data. 
The expected
effect should be a few km~s$^{-1}$, but it probably drowns in the noise
from spectroscopic measurements that have errors of comparable magnitude,
and from a possible anisotropic expansion.

The interpretation of Fig.~\ref{fig7} is not obvious.  Stars in Lower
Centaurus Crux show a wide spread in the spectroscopic values, while the
mean is roughly consistent with an isotropic expansion at about half the
rate naively expected from the age of the association. The same can be said for
Upper Centaurus Lupus. For these associations the indicated `kinematic
age' (equal to the inverse of the current expansion rate) is thus around
20--30~Myr, or twice the isochrone ages according to de Geus et al.\ (\cite{degeus}).
Upper Scorpius on the other hand, which is the youngest of the subgroups 
(5--6~Myr according to de Geus et al.), does not seem to
expand at all: taken at face value, the data rather suggest that it contracts. 
The combined sample again indicates \textit{some} expansion,
roughly consistent with a kinematic age of 20~Myr.
We note that already Blaauw (\cite{blaauw}) derived such an expansion age of 20~Myr
for the Scorpius-Centaurus complex as a whole from the $\simeq 10$~km~s$^{-1}$
discrepancy between the spectroscopic radial velocities and the proper
motion data combined with photometric distances.

Of the three subgroups in Sco~OB2, the result for Upper Scorpius thus 
stands out as rather puzzling. A detailed study of this
association by Preibisch \& Zinnecker (\cite{preibisch}) suggested that the star 
formation process was triggered by a giant supernova explosion in the 
neighbouring Upper Centaurus Lupus. What effect that may have had on 
the internal kinematics of Upper Scorpius is hard to say. There is 
a~priori no reason to expect Upper Scorpius to be a bound system without
expansion. The star formation in Upper Scorpius itself seems to have
dispersed the rest of the parent molecular cloud. This result seems to
imply the standard picture (see e.g.\ Mathieu \cite{mathieu} for a review): the
removal of gas leads to loss of binding mass of the system, it becomes
unbound and consequently will expand.

From calculations inspired by our method, Makarov \& Fabricius (\cite{makarov01})
estimated an expansion rate of
0.12~km~s$^{-1}$ pc$^{-1}$ for the TW~Hya association of young
stars, assuming a uniform expansion.
The TW~Hya association is dominated by late-type stars
and may be an extension of Lower Centaurus Crux. The expansion
corresponds to a bias of the centroid radial velocity of  
$\sim -9$~km~s$^{-1}$ -- comparable to the biases we find for 
Upper Centaurus Lupus and Lower Centaurus Crux -- and to a dynamical 
age of 8.3~Myr, in agreement with previous age determinations for 
TW~Hya's T~Tauri members (Webb et al.\ \cite{webb}).

The results we find here are promising in the sense that it is possible
to obtain information about the internal kinematics, formation history
and age, but at the same time they confirm the complexity of the kinematics 
of the associations in the Sco~OB2 complex. In the end more accurate 
spectroscopic observations are also required to answer these questions.
These would in particular allow true expansion to be disentangled from 
the perspective effects of the radial motion.

\section{Conclusions}
\label{sec:conclus}

The radial motions of stars have been studied through spectroscopy since the year
1868 (Hearnshaw \cite{hearnshaw}).  Recently, the accuracies realized in astrometry have 
enabled such determinations to be made also through purely geometric methods.  
Once sufficient accuracies are reached, this will enable an absolute calibration 
of the stellar velocity scale for stationary and variable stars, irrespective of 
any complexities in their spectra. 
Indeed, for several early-type stars (with complex spectra smeared by their
rapid rotation) the radial velocities already now determined through
astrometry are more accurate than has been possible to reach
spectroscopically in the past.

 The differences between these astrometric 
radial-velocity values and wavelength measurements of different spectral 
features may become a new diagnostic tool in probing the dynamic structure of 
stellar atmospheres.
Already the present work has made available quite accurate astrometric
radial velocities for stars of many more spectral types than those for which
hydrodynamic model atmospheres have been developed (from which, e.g.,
convective and gravitational wavelength shifts in their spectra could have
been predicted).  For such stars, the limitations in understanding the
differences between astrometric and spectroscopic radial velocities may now
lie primarily with spectroscopy and atmospheric modelling, rather than in
astrometry.

In this series of three papers, we started by exploring different types of 
fundamental possibilities of astrometrically determining radial velocities, 
identifying which methods could be applicable on existing data already today.  
Among the latter, the moving-cluster method was found capable of yielding 
astrophysically interesting, sub-km~s$^{-1}$ accuracies, and its mathematical 
methods were developed in Paper~II.  In the present paper, data from Hipparcos 
were used in applying the method to obtain solutions for more than 1\,000 
stars in nearby clusters and associations.  Although most of these do not reach 
the high accuracies realized for the Hyades, they hint at the future potential.

Quantitatively, we have obtained radial velocities with standard errors of 
$\sim$ 0.6~km~s$^{-1}$ for individual stars in the Hyades. 
The accuracies reached begin to make visible the convective and gravitational 
shifts expected in the spectra of F and G stars. For A stars and
earlier types, where the convective shifts cannot yet be reliably
predicted from theory, the spectra appear to be blueshifted by a 
few km~s$^{-1}$ compared with the astrometrically determined motions 
and expected gravitational redshifts.
This illustrates that astrometric radial velocities with
uncertainties even well in excess of 1~km~s$^{-1}$ could be
astrophysically interesting. 

Such accuracies may also be sufficient to provide information about the 
expansion of OB associations, as illustrated by the results for the 
Sco~OB2 complex. Even with the modest precision of existing spectroscopic
velocities, we see indications of expansion in the OB associations 
Upper Centaurus Lupus and Lower Centaurus Crux (causing a bias in the
astrometric radial velocities of 5--10~km~s$^{-1}$), while Upper Scorpius
surprisingly shows no such indication. The limitations in the present understanding 
of these associations come not from astrometry but mainly from spectroscopy 
and theory.

From the same solution that gave astrometric radial velocities, we get
kinematically improved parallaxes. These can be used to study in greater
detail the spatial structures and the Hertzsprung-Russell diagrams of both 
clusters and OB associations. 
In case of the Hyades, Upper Centaurus Lupus and Lower Centaurus Crux,
the better-defined main sequences can also be taken as proof of the
validity of the kinematic solution, and hence of the astrometric 
radial velocities.

Hipparcos parallax measurements reached typical accuracies of about 
1.5~mas, while our improved parallaxes reach 0.3~mas for the Hyades. Space 
astrometry missions in the near future are expected to improve this by more than 
an order of magnitude to about 0.05~mas (Horner et al.\ \cite{horner}), with another 
order-of-magnitude gain by the future GAIA to 0.004~mas (Perryman et al.\
\cite{perryman01}).  As detailed in Paper~I, such accuracies will enable also other methods 
than the moving-cluster one for determining radial velocities by purely 
geometric means.  The future prospects for studying absolute radial velocities 
independent of spectroscopy look exciting indeed!

\begin{acknowledgements}

This project is supported by the Swedish National Space Board and the Swedish 
Natural Science Research Council.  We want to thank Tim de Zeeuw for providing 
data on several nearby OB associations before publication, Floor van Leeuwen 
for providing data on the Pleiades and Praesepe clusters, and the referee,
Anthony Brown, for valuable comments.

\end{acknowledgements}

\end{document}